\documentclass[twocolumn]{aastex631}

\usepackage{graphicx}
\usepackage{longtable}
\usepackage{amsmath}
\usepackage{booktabs}
\usepackage{multirow}
\usepackage{float}

\accepted{April 26, 2022}

\submitjournal{ApJ}

\shorttitle{Outflows in Bulgeless Galaxies}
\shortauthors{Bohn et al.}

\graphicspath{{./}{figures/}}

\begin{document}

\title{A Near-Infrared look at AGN Feedback in Bulgeless Galaxies}

\email{tbohn002@ucr.edu}

\author[0000-0002-4375-254X]{Thomas Bohn}
\affil{University of California, Riverside, Department of Physics \& Astronomy \\
900 University Ave., Riverside, CA 92521}

\author[0000-0003-4693-6157]{Gabriela Canalizo}
\affiliation{University of California, Riverside, Department of Physics \& Astronomy \\
900 University Ave., Riverside, CA 92521}

\author[0000-0003-2277-2354]{Shobita Satyapal}
\affiliation{George Mason University, Department of Physics \& Astronomy \\
MS3F3, 4400 University Drive, Fairfax, VA 22030}

\author[0000-0002-3790-720X]{Laura V. Sales}
\affiliation{University of California, Riverside, Department of Physics \& Astronomy \\
900 University Ave., Riverside, CA 92521}

\begin{abstract}

While it is generally believed that supermassive black holes (SMBH) lie in most galaxies with bulges, few SMBHs have been confirmed in bulgeless galaxies. Identifying such a population could provide important insights to the BH seed population and secular BH growth. To this end, we obtained near-infrared spectroscopic observations of a sample of low-redshift bulgeless galaxies with mid-infrared colors suggestive of AGN. We find additional evidence of AGN activity (such as coronal lines and broad permitted lines) in 69$\%$ (9/13) of the sample, demonstrating that mid-infrared selection is a powerful tool to detect AGN. More than half of the galaxies with confirmed AGN activity show fast outflows in [\ion{O}{3}] in the optical and/or [\ion{Si}{6}] in the NIR, with the latter generally having much faster velocities that are also correlated to their spatial extent. We are also able to obtain virial BH masses for some targets and find they fall within the scatter of other late-type galaxies in the $M_{\rm{BH}}$-$M_{\rm{stellar}}$ relation. The fact that they lack a significant bulge component indicates that secular processes, likely independent of major mergers, grew these BHs to supermassive sizes. Finally, we analyze the rotational gas kinematics and find two notable exceptions: two AGN hosts with outflows that appear to be rotating faster than expected. There is an indication that these two galaxies have stellar masses significantly lower than expected from their dark matter halo masses. This, combined with the observed AGN activity and strong gas outflows may be evidence of the effects of AGN feedback.

\end{abstract}

\keywords{galaxies: active --- galaxies: bulges --- galaxies: evolution --- galaxies: Seyfert --- infrared: galaxies}

\section{Introduction} \label{sec:intro}

The relations between the properties of galaxies and their SMBHs have become a vital part of understanding galaxy evolution. Well-known correlations such as the $M_{\rm{BH}}$-$\sigma$ \citep[eg.,][]{Ferrarese2000,Gebhardt2000,McConnell2013,Sexton2019} and the $M_{\rm{BH}}$ - $M_{\rm{bulge}}$, \citep[eg.,][]{Marconi2003,Haring2004} have shown that SMBHs and their host galaxies likely co-evolve with each other. In the larger mass regime, galaxy mergers are often regarded as the driving mechanism behind this co-evolution. This is because galaxy mergers are able to disrupt the angular momentum of the baryonic matter rotating in the disk. This redistribution culminates not only in a galactic bulge forming but can also result in gas being funneled toward the central BH. It is through these processes that BHs are believed to reach supermassive sizes. 

This paradigm of BH growth accompanying bulge growth raises an important question regarding galaxies that have not had merger events in their history. While simulations have shown that galaxies can regrow their disks after a merger event \citep[eg.,][]{Hopkins2009}, a small bulge component can still form after a minor merger. Thus, it is likely that a galaxy with no bulge has had a merger-free history. These systems are of particular interest since their BHs have grown through secular processes. As such, their number density and occupation fraction can provide important constraints on the BH seed population. Indeed, a number of studies have investigated BHs residing in bulgeless host galaxies. \citet{Simmons2017} report BH mass estimates for a large sample of Type 1 targets and find they fall within the scatter of the $M_{\rm{BH}}$ - $M_{\rm{stellar}}$ relation. Likewise, \citet{Bohn2020} come to a similar conclusion for a Type 2 AGN through the use of near-infrared (NIR) observations. Adding to this fact that some simulations suggest mergers do not induce a significant amount of SMBH growth \citep{Martin2018,McAlpine2020}, secular processes may play a more significant role in BH growth than originally thought.

The addition of SMBHs in bulgeless galaxies necessitates a closer look at these systems. Understanding the dynamics between these BHs and their hosts is crucial to putting constraints on secular growth processes and BH-galaxy co-evolution. An important aspect of this co-evolution is the impact of AGN feedback. Outflows from AGN are believed to regulate and suppress star formation, leading their hosts to the well-defined red sequence \citep{King2015,Rupke2017}. However, only a limited number of AGN-driven outflows have been observed in bulgeless galaxies \citep{Smethurst2019,Smethurst2021}. In order to place more stringent constraints and assess the impact of secularly grown BHs, a larger sample size is needed.

A central issue when identifying AGN in bulgeless galaxies is confirming the true morphology of the host. While estimates to $M_{\rm{BH}}$ may be relatively easy to obtain in Type 1 AGN through the use of emission from the broad-line region and the virial method, the reliability of bulge-disk decompositions could be compromised due to the bright AGN being in our direct line of sight. Consequently, a small bulge component could be hidden. Type 2 AGN benefit from more robust bulge-disk decompositions but the lack of broad emission lines, particularly in optical wavelengths, makes it difficult to obtain robust $M_{\rm{BH}}$ estimates. An added limitation is that bulgeless galaxies can have high levels of dust obscuration, making AGN identification difficult and thus optical selection incomplete \citep{Sat14}.

A number of methods have been suggested to identify and confirm the presence of obscured AGN, including spectropolarimetry \citep{Antonucci1985,Zakamska2005} and infrared observations \citep[eg.,][]{Veilleux1997,Satyapal2007,Satyapal2008,Sat14,Lamperti2017,Cann2020}, where extinction is less severe in the latter. The NIR spectral region hosts a number of emission lines used for AGN analysis. Below z $\sim$ 0.3, these include coronal lines (CLs), forbidden transitions from highly ionized ($>$100 eV) species that are highly indicative of AGN activity, and \ion{H}{1} recombination lines that can trace the broad-line gas. Additionally, due to the characteristically redder mid-infrared (MIR) colors of obscured AGN \citep{Stern2012}, MIR color cuts can be used to separate optically hidden AGN from star forming galaxies. Indeed, the \textit{Wide-field Infrared Survey Explorer} (\textit{WISE}; \citealt{Wright2010}) and \textit{Spitzer Space Telescope} have been used to separate AGN from star-forming galaxies, as shown by the AGN demarcation box in \citet{Jarrett2011}. Recent follow-up work has indicated that this color-cut could identify a new group of obscured AGN that would complement existing optical data and provide a more representative picture of the AGN population \citep[][hereafter S14]{Sat14}. However, confirmation that these galaxies do indeed host AGN is needed since contamination from star formation cannot be dismissed.

To this end, in this article we aim to confirm the presence of AGN activity in the sample of MIR-selected bulgeless galaxies presented in S14. In Section \ref{sec:Observations}, we provide details of the sample, observations, and data reduction. Fitting of the spectra and calculations of the extinction levels and star-formation rates (SFRs) are covered in Section \ref{sec:Analysis}. In Section \ref{sec:AGN_indicators}, we use a number of AGN indicators, including CLs, broad-line emission, and AGN diagnostic diagrams, to confirm the presence of AGN activity. Section \ref{sec:outflows} analyzes outflow kinematics and energies. We report virial BH mass estimates in Section \ref{sec:BH_Comparisons} and compare these to known scaling relations. Lastly, in Section \ref{sec:Gas_Kinematics}, we analyze rotational gas velocities and discuss the affects of AGN feedback in our sample. We adopt a standard $\Lambda$CDM cosmology with $H_0$ = 70 km s$^{-1}$ Mpc$^{-1}$, $\Omega_M$ = 0.3, and $\Omega_\Lambda$ = 0.7.

\section{Data and Observations} \label{sec:Observations}

\begin{figure*}
\centering
\epsscale{1.2}
\plotone{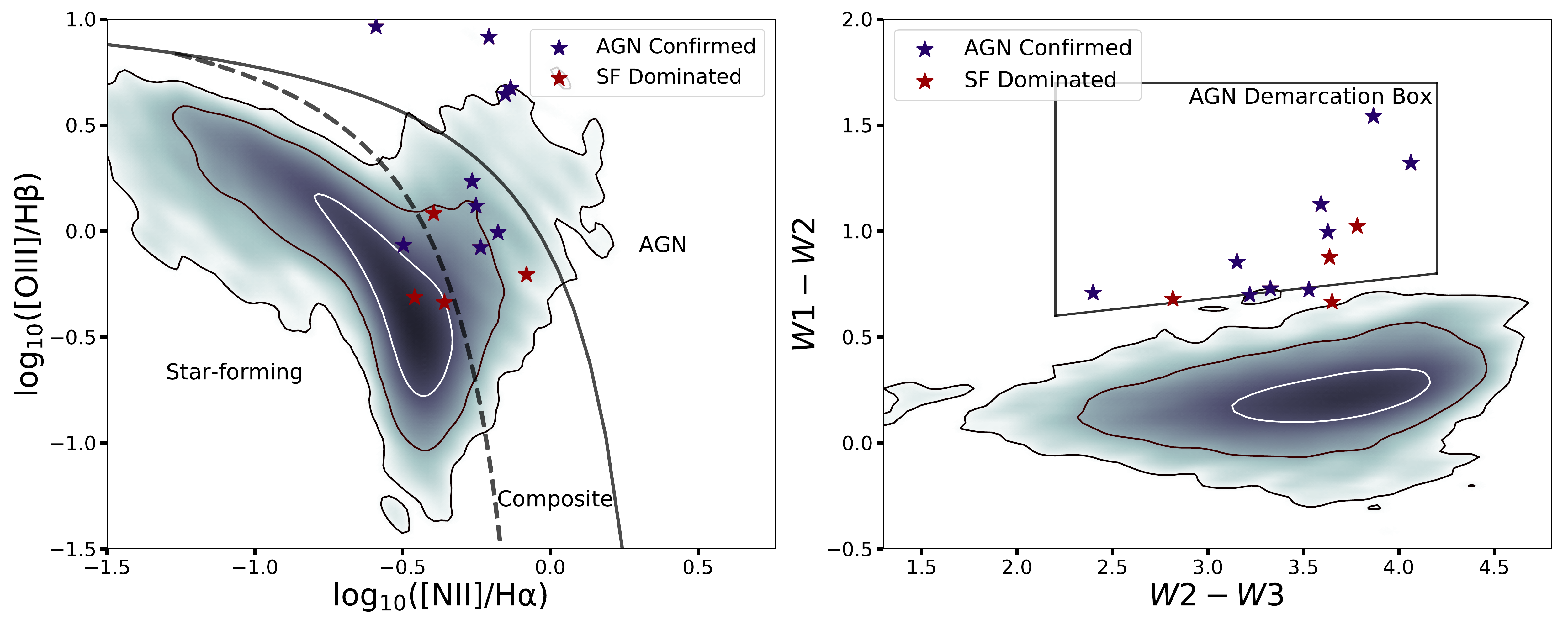}
\caption{(\textit{left}) \textit{BPT} diagram with diagnostic lines from \citet{Kewley2001} and \citet{Kauffmann2003}. The full bulgeless sample from \citet{Simard2011} are represented as contours (1$\sigma$, 2$\sigma$, and 3$\sigma$ levels). Galaxies covered in this study are represented as stars, with those confirmed as AGN with at least one indicator are shown as blue stars (see Section \ref{sec:AGN_indicators}). Targets whose emission is dominated by stellar processes are shown red stars. (\textit{right}) \textit{WISE} color diagram with the AGN demarcation box from \citet{Jarrett2011}. Emission line fluxes are provided in Table \ref{tab:table_fluxes}. \label{fig:WISE}}
\end{figure*}

\subsection{Sample Selection} \label{subsec:Selection}

Our sample of bulgeless galaxies come from those presented in S14. To summarize their selection process, they drew from \citet{Simard2011}, where bulge-disk decompositions were performed on SDSS imaging. To obtain a bulgeless subsample, S14 selected galaxies with a bulge-to-total light ratio of 0.00 in both $r$ and $g$-bands. SDSS spectroscopy was also required for proper redshift estimates, which resulted in a bulgeless sample of $\sim$18,000 galaxies (see Figure \ref{fig:WISE}). They subsequently cross-matched this bulgeless subsample with the \textit{WISE} all-sky data release catalog\footnote{\url{https://wise2.ipac.caltech.edu/docs/release/allsky/}} in order to obtain $W1$ (3.4 $\mu$m), $W2$ (4.6 $\mu$m), and $W3$ (12 $\mu$m) bands. Utilizing the AGN demarcation zone as defined in \citet{Jarrett2011}, they then formed a sample of 30 AGN candidates, all of which have $W1$ -- $W2$ $>$ 0.7. These candidates form the basis of our sample. In total, we obtained NIR spectroscopy of 13 of these 30 targets during our assigned telescope time, and these 13 targets span the full range of \textit{WISE} colors presented in S14. In Figure \ref{fig:WISE}, we plot these targets on the Baldwin, Phillips $\&$ Terlevich \citep[BPT;][]{Baldwin1981} diagram and the \textit{WISE} color-cut diagram as presented in \citet{Jarrett2011}. Hereafter, targets with at least one AGN indicator will be represented as a blue symbol while those whose emission is dominated by stellar processes will be shown as red symbols (see Section \ref{sec:AGN_indicators} for details). The redshift range of our sample is from 0.05 to 0.24, with a median of 0.117. The stellar masses range from 10.0 - 11.1 M$_\odot$ and were obtained from \citet{Chang2015}, who performed SED fitting on optical and IR data. Comparing to the bulgeless sample as a whole, our galaxies tend to be on the higher mass end. 

\subsection{Observations and Data Reduction} \label{subsec:Observe}

Our Keck NIR spectroscopy was obtained throughout the 2018 and 2019 semesters using two instruments: NIRSPEC \citep{McLean1998} and NIRES \citep{Wilson2004}. NIRSPEC is a NIR echelle spectrograph on Keck II with a wavelength coverage of 0.9 --- 5.5 $\mu$m. The NIRSPEC-7 filter was used in low-resolution mode ($42''\times0.76''$ slit) and the cross-dispersion angle was set to 35.38$^{\circ}$. This resulted in a wavelength coverage of 1.98 --- 2.40 $\mu$m. An average spectral resolution of 197 km s$^{-1}$ ($R$ $\approx$ 1500) was measured at 2.20 $\mu$m. All NIRSPEC observations were done on 2018-03-05 under mostly clear conditions and the seeing was typically $\sim$0.6$''$.

NIRES is a NIR echelette spectrograph with a 18$''\times0.55''$ slit and wavelength coverage of 0.94 --- 2.45 $\mu$m, split into five orders. There is a small gap in coverage between 1.85 and 1.88 $\mu$m, but this is a region of low atmospheric transmission. The average spectral resolution of the five orders ranges between 85 --- 95 km s$^{-1}$ (R $\approx$ 3300). The majority of observations were done under mostly clear conditions ($\sim$0.5" seeing); only J0122+0100 and J0333+0107 were observed under poor weather. 

Individual exposures for all sets of observations were four minutes each and were done using the standard ABBA nodding. A telluric standard star, typically of A0 spectral class with measured magnitudes in $J$, $H$, and $K$-bands, was observed either directly before or after the target galaxy to correct for atmospheric absorption features. Typical airmass differences with the target were below 0.10. Details of the observations are summarized in Table \ref{tab:obs_log} and PanSTARR \citep{Chambers2016} images of our sample are shown in Figure \ref{fig:Slit_pos}. For J1036+0221 we provide \textit{HST} imaging. Slit positions and contours showing the isophotes are also provided.

\begin{deluxetable*}{ccccccccc}
\caption{Observation Log} 
\label{tab:obs_log}
\tablehead{\colhead{Galaxy} & \colhead{Date} & \colhead{Instrument} & \colhead{Redshift} & \colhead{Scale} & \colhead{Exp. Time\tablenotemark{a}} & \colhead{Slit PA} & \colhead{Aperture} & \colhead{Airmass} \\
\colhead{(SDSS Name)} & \colhead{(YYYY-mm-dd)} & \colhead{} & \colhead{} & \colhead{(kpc/arcsec)} & \colhead{} & \colhead{(degrees)} & \colhead{(arcsec)} & \colhead{}}
\startdata
J012218.11+010025.76 & 2018-09-20 & NIRES & 0.05546 & 1.08 & 4 x 240s\tablenotemark{b} & 100, 183 & 1.92 & 1.25\\
J033331.86+010716.92 & 2018-10-24 & NIRES & 0.17755 & 3.00 & 2 x 240s & 0 & 1.49 & 1.15\\
J075039.97+292845.75 & 2018-03-05 & NIRSPEC & 0.14701 & 2.57 & 12 x 240s & 68 & 1.46 & 1.11\\
J085153.64+392611.76\tablenotemark{c} & 2019-03-25 & NIRES & 0.12958 & 2.31 & 5 x 240s & 94 & 1.33 & 1.07\\
--- & 2018-03-05 & NIRSPEC & --- & --- & 8 x 240s & 46 & 1.34 & 1.06\\
J085236.35+292853.20 & 2019-03-25 & NIRES & 0.07866 & 1.49 & 8 x 240s & 162 & 1.34 & 1.04\\
J092907.78+002637.29 & 2019-03-25 & NIRES & 0.11732 & 2.12 & 8 x 240s & 56.5 & 1.51 & 1.09\\
J103222.94+361727.91 & 2019-03-25 & NIRES & 0.23578 & 3.74 & 4 x 240s & 131 & 1.18 & 1.05\\
J103631.88+022144.10 & 2018-03-05 & NIRSPEC & 0.05026 & 0.98 & 4 x 240s & 42.5 & 1.68 & 1.05\\
J122809.19+581431.40\tablenotemark{c} & 2019-01-24 & NIRES & 0.10982 & 2.00 & 12 x 240s & 333 & 1.34 & 1.28\\
--- & 2018-03-05 & NIRSPEC & --- & --- & 8 x 240s & 333 & 1.34 & 1.27\\
J123304.57+002347.11 & 2019-03-25 & NIRES & 0.06917 & 1.32 & 8 x 240s & 75.5 & 1.33 & 1.06\\
J132647.54+101436.53 & 2018-03-05 & NIRSPEC & 0.08943 & 1.67 & 6 x 240s & 161.5 & 1.57 & 1.02\\
J133341.72+653619.16 & 2019-03-25 & NIRES & 0.15756 & 2.72 & 4 x 240s\tablenotemark{d} & 123, 44 & 1.34 & 1.43\\
J155409.08+145703.59 & 2019-03-25 & NIRES & 0.13666 & 2.42 & 8 x 240s & 121 & 1.33 & 1.02
\enddata
\tablenotetext{a}{Exposures were typically done in ABBA nodding.}
\tablenotetext{b}{Four exposures were taken for both the east and west targets.}
\tablenotetext{c}{J0851 and J1228 were also observed with NIRSPEC.}
\tablenotetext{d}{Four exposures were taken for each position angle.}
\end{deluxetable*}
\vspace*{-8mm}

A similar reduction process was performed for both the NIRSPEC and NIRES data. However, the differences in the two instruments necessitated slight modifications in the two pipelines used. The first algorithm provided flat fielding and a robust background subtraction by using techniques described in \citet{Kelson2003} and \citet{Becker2009}. In short, this routine maps the 2D science frame and models the sky background before rectification, thus reducing the possibility of artifacts appearing due to the binning of sharp features. The sky subtraction attained with this procedure is excellent, despite the strong OH lines present in the NIR; the procedure is also quite insensitive to cosmic rays and hot pixels, and is reliable regardless of skyline intensity.

Rectification, telluric correction, wavelength calibration, and extraction were all done with a slightly modified version of \textsc{REDSPEC}.\footnote{\url{https://www2.keck.hawaii.edu/inst/nirspec/redspec.html}} Telluric correction was done by dividing by the spectrum of the telluric standard star and multiplying by a blackbody curve of the same temperature. Strong OH skylines were used for wavelength calibration and the 1D spectra were then median combined. Flux calibration of individual exposures was done using the telluric star and the Spitzer Science Center unit converter\footnote{\url{https://irsa.ipac.caltech.edu/data/SPITZER/docs/dataanalysistools/tools/pet/magtojy/index.html}} to convert the magnitude of the star to the associated flux in that band. In addition to correcting for Galactic reddening, a small corrective factor ($<$5\%) was introduced due to the differences between the center NIR bands and that of the wavelength coverage.

\section{Analysis} \label{sec:Analysis}

\subsection{Spectral Fitting} \label{subsec:Fitting}

\subsubsection{NIR Fitting} \label{subsubsec:NIR_fit}

\begin{figure*}
\centering
\epsscale{1.0}
\plotone{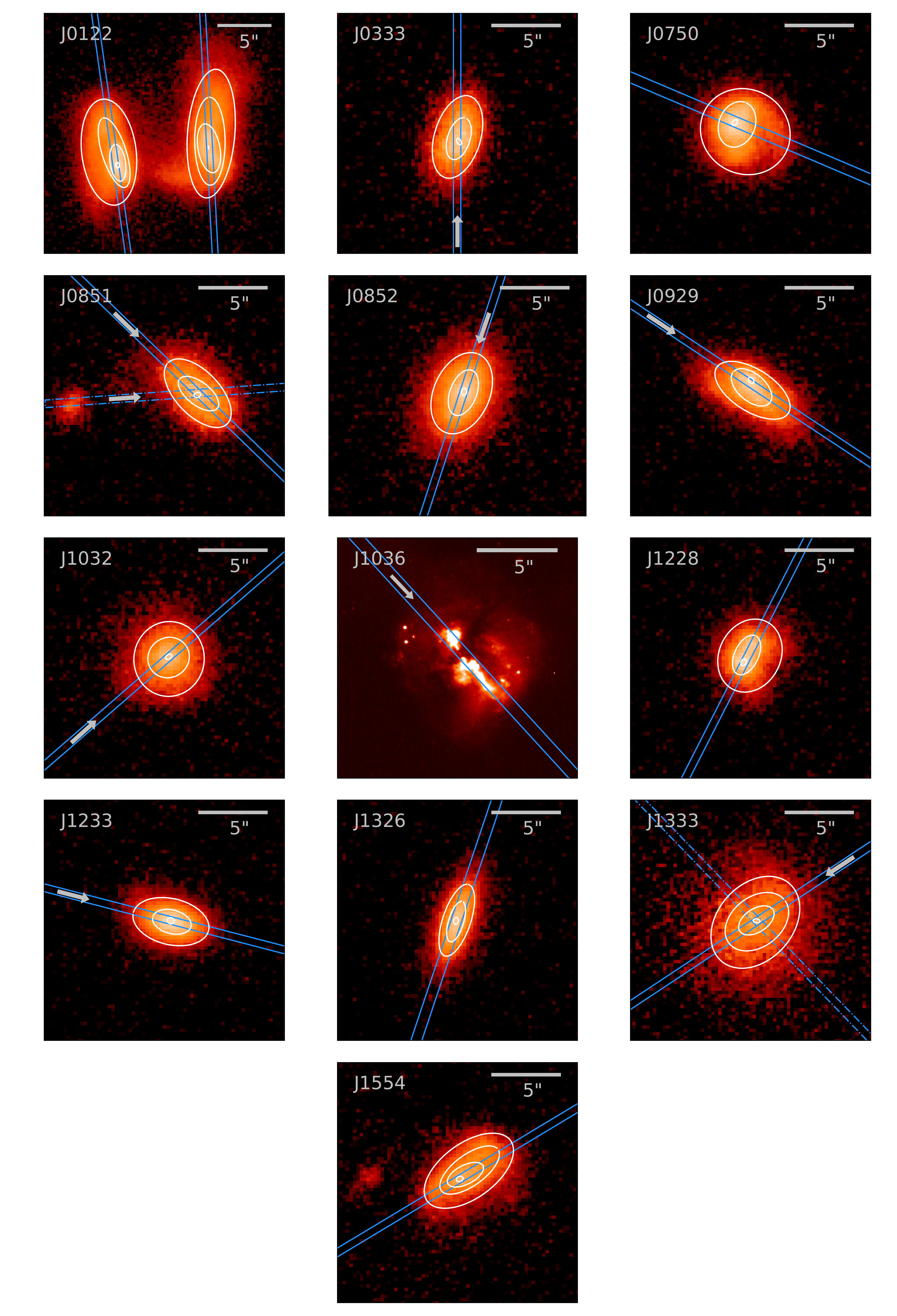}
\caption{PanSTARRS (and \textit{HST} for J1036+0221) image cut-outs of our sample. The galaxy short-name and 5" scale are listed at the top. The blue lines indicate the placement of the slit. For galaxies where we see gas rotation, we place arrows pointing in the direction from blueshifted to redshifted velocities (see Fig. \ref{fig:rotation_curve}). Lastly, the white contours show the isophotes of each target. \label{fig:Slit_pos}}
\end{figure*}

We fit all NIR spectra using \textsc{emcee}, an affine invariant Markov Chain Monte Carlo (MCMC) ensemble sampler \citep{Foreman-Mackey2013}. A narrow Gaussian component, along with a low order ($\leq$ 7) Legendre polynomial for the continuum, were fit simultaneously for each emission line. We determined whether a second Gaussian component was needed to fit the emission lines using the following \textit{F}-test: \textit{F} = $(\sigma_{single})^{2}/(\sigma_{double})^{2}$, where $\sigma$ is the standard deviation of the residuals using either single or double Gaussian components. If \textit{F} $>$ 2.0, then adding an extra component is justifiable and we added a component that is constrained to be broader and lower in amplitude than the first component to avoid degeneracies. Both the narrow and broad components were treated as free Gaussians with amplitude, full-width half-maximum (FWHM), and velocity offset from rest-frame wavelength as free variables. The fluxes of all NIR emission lines detected at $>$ 3$\sigma$ level are listed in Table \ref{tab:table_fluxes} of Appendix \ref{appendix:line_fluxes}.

Various absorption features were also fit in a similar fashion. The fitting was done simultaneously with that of emission lines in order to keep the continuum level consistent across all measurements. Although we ran the \textit{F}-test as defined above, only one gaussian was needed for all the absorption fits.

\subsubsection{SDSS Fitting} \label{subsubsec:SDSS_fit}

As noted in Section \ref{subsec:Selection}, SDSS spectra are available for our entire sample and these provide full wavelength coverage from 4000 \AA\; to 9000 \AA. The SDSS spectra were fit using Bayesian AGN Decomposition Analysis for SDSS Spectra \citep[{\textsc{BADASS,}\footnote{\url{https://github.com/remingtonsexton/BADASS2}}}][]{Sexton2020}, a spectral analysis tool that provides comprehensive fits to SDSS spectra. Through \textsc{BADASS}, we fit absorption features using the penalized Pixel Fitting \citep[\textsc{pPXF}\footnote{\url{https://www-astro.physics.ox.ac.uk/~mxc/software/}};][]{Cappellari2004} with stellar templates from The Indo-US Library of Coud{\'e} Feed Stellar Spectra \citep{Valdes2004}. We also included a power-law function to simulate the AGN continuum. However since our sample does not have Type 1 Seyferts, we did not fit for FeII emission. All components were fit simultaneously, allowing for a detailed and robust analysis of the spectrum. Notable emission line fluxes are listed in Table \ref{tab:table_fluxes}.

A key feature of \textsc{BADASS} is that it allows the user to test for the presence of outflows in the form of broad components of emission lines by setting various constraints on parameters such as amplitude, line width, and velocity offset. [\ion{O}{3}] $\lambda$5007 is often used as a tracer of ionized gas outflows in the optical since it is isolated (i.e., no risk of blending with broad-lines) and in a region free of significant stellar absorption. Evidence of outflow activity is generally believed to manifest itself as a broad flux excess, typically seen as an asymmetric wing that is blue-shifted relative to the rest-wavelength emission \citep{Woo2016,Manzano2019}. In addition to velocity shifts, widths of a few hundred to thousands of km s$^{-1}$ have been reported in optical and NIR data \citep[eg.,][]{Rodriguez2006,Muller2011,Harrison2014,McElroy2015}. These fast outflows can be seen on kiloparsec (kpc) scales \citep[eg.,][]{Cresci2015,Rupke2017,Muller2018a,Liu2020} and could potentially be key contributors to suppressing star formation \citep{Fabian2012,King2015,Costa2020}. If the outflows are driven by the AGN, then this BH feedback may lead to well-known scaling relations, such as the $M_{\rm{BH}}$-$\sigma$. In addition, CL emission has often been observed with strong outflows \citep{Muller2011,Bohn2021}, making outflow detections a unique point of interest when searching for and confirming AGN activity using CLs.

Through SDSS spectral fitting of [\ion{O}{3}], we detect strong outflows in four galaxies: J0852+2928, J0929+0026, J1233+0023, and J1333+6536 at a $>$ 95$\%$ confidence based on the \textit{F}-test model comparison included in the code. We also require the amplitude of the outflow component to be greater than 3$\sigma$ of the noise level and FWHM to be 1$\sigma$ greater than the core component. This confidence cut and parameter requirements result in a stringent selection criteria. In the successful cases, the outflow parameters determined in [\ion{O}{3}] are used to constrain the outflow components found in other strong emission lines (H$\alpha$, H$\beta$, etc.). These constraints include a fixed amplitude ratio ($A_{outflow}/A_{core}$), FWHM, and velocity offset. This is done since blending, particularly between H$\alpha$ and [\ion{N}{2}], can be problematic in identifying and fitting outflow components. We have listed fluxes to the outflow components in Table \ref{tab:table_fluxes}.

A non-outflow detection worthy of note is J1036+0221 since it does show a high confidence, $\sim$90$\%$. However, it is likely undergoing a merger event and a number of mechanisms could be causing the broader [\ion{O}{3}] profile, such as superposition of two nuclei or tidal disruptions caused by the merger. Due to these uncertainties, we do not include J1036+0221 in our outflow analysis.

We do not detect outflows in the rest of our sample, where they all have low confidence levels ($<$ 65$\%$). However, this does not rule out the presence of outflows in these galaxies. As \citet{Sexton2020} points out, signal-to-noise (S/N) can play an important role in detecting outflows. Additionally, spatial and spectral resolution can affect detection, as shown in \citet{Manzano2019}, where nine outflows are detected in Keck LRIS but only two of these are detected in SDSS. Kadir et al. (\textit{in prep.}) also find that high extinction can inhibit outflow detection.  Indeed, galaxies in our sample where we do see outflows tend to have lower levels of extinction (see Tables \ref{tab:relations} and \ref{tab:outflow_kinematics}). Thus the detection rate of outflows in our sample should be taken as a lower limit.

\begin{deluxetable*}{ccccccccccc}
\caption{Summary of Galaxy Properties} 
\label{tab:relations}
\tablehead{\colhead{Galaxy} & \colhead{W1-W2} & \colhead{log(M$_{BH}$)} & \colhead{E(B-V)} & \colhead{SFR} & \colhead{CL?} & \colhead{Outflow?} & \colhead{Overmass.} & \colhead{BPT} & \colhead{OVT '92} & \colhead{AGN}\\
\colhead{ } & \colhead{ } & \colhead{($M_\odot$)} & \colhead{ } & \colhead{($M_\odot$ yr$^{-1}$)} & \colhead{ } & \colhead{ } & \colhead{DMH?} & \colhead{AGN?} & \colhead{AGN?} & \colhead{Evidence?}\\
\colhead{(1)} & \colhead{(2)} & \colhead{(3)} & \colhead{(4)} & \colhead{(5)} & \colhead{(6)} & \colhead{(7)} & \colhead{(8)} & \colhead{(9)} & \colhead{(10)} & \colhead{(11)}}
\startdata
J0122ne & ---  & --- & 0.00 & $<$0.40 & --- & --- & --- & --- & \checkmark & \checkmark\\
J0122se & 1.60  & --- & 0.80$^{+0.11}_{-0.11}$ & $<$3.47 & \checkmark\tablenotemark{a} & --- & --- & --- & --- & \checkmark\\
J0122w & --- & --- & 1.82$^{+0.07}_{-0.07}$ & $<$8.57 & --- & --- & --- & --- & --- & ---\\
J0333 & 0.87 & --- & 1.38$^{+0.28}_{-0.31}$ & 4.00$^{+2.16}_{-1.50}$ & --- & --- & --- & --- & --- & ---\\
J0750 & 1.00 & --- & 1.27$^{+0.49}_{-0.57}$ & 3.28$^{+3.59}_{-1.91}$ & --- & --- & --- & --- & --- & ---\\
J0851 & 0.71 & 6.78$\pm$0.14 & 0.66$^{+0.36}_{-0.40}$ & 0.36$^{+0.32}_{-0.19}$ & --- & --- & --- & --- & \checkmark & \checkmark\\
J0852 & 0.76 & --- & 0.00 & $<$0.18 & \checkmark & \checkmark & \checkmark & \checkmark & \checkmark & \checkmark\\
J0929 & 0.85 & --- & 1.28$^{+0.08}_{-0.08}$ & --- & \checkmark* & \checkmark & --- & \checkmark & \checkmark & \checkmark\\
J1032 & 0.74 & --- & 0.80$^{+0.25}_{-0.27}$ & 2.74$^{+1.33}_{-0.96}$ & --- & --- & --- & --- & --- & ---\\
J1036 & 1.30 & --- & 2.17$^{+0.13}_{-0.13}$ & $<$6.53 & \checkmark\tablenotemark{a} & --- & --- & --- & \checkmark & \checkmark\\
J1228 & 0.72 & 6.80$\pm$0.32 & 2.50$^{+1.25}_{-2.12}$ & $<$3.19 & --- & --- & --- & \checkmark & \checkmark & \checkmark\\
J1233 & 0.98 & --- & 0.81$^{+0.30}_{-0.34}$ & $<$0.46 & \checkmark & \checkmark & \checkmark & --- & --- & \checkmark\\
J1326 & 0.73 & --- & 2.88$^{+1.08}_{-1.67}$ & $<$2.04 & --- & --- & --- & --- & --- & ---\\
J1333 & 0.73 & --- & 0.18$^{+0.33}_{-0.18}$ & --- & \checkmark* & \checkmark & --- & \checkmark & \checkmark & \checkmark\\
J1554 & 1.14 & --- & 1.59$^{+0.47}_{-0.55}$ & 1.18$^{+1.69}_{-0.78}$ & \checkmark & \checkmark & --- & --- & --- & \checkmark
\enddata
\tablecomments{Columns: (1) Galaxy name. (2) \textit{WISE} \textit{W1}-\textit{W2} magnitude. (3) BH mass as determined in Section \ref{subsec:BH_mass} and \ref{sec:BH_Comparisons}. (4) Extinction based on the H$\alpha$/H$\beta$ Balmer decrement. (5) Star-formation rate based on [\ion{O}{2}] \citep{Kewley2004}. We omit values if [\ion{Ne}{5}] is detected and place upper limits if the presence of [\ion{Ne}{5}] cannot be determined. (6) Presence of NIR CLs. An asterisk (*) signifies optical CLs are detected. (7) Evidence of outflows. (8) Evidence of an overmassive DMH. (9) Galaxy falls within the AGN region of the BPT diagram. (10) Galaxy falls within the AGN region of at least one of the diagnostic diagrams in \citet{Osterbrock1992}. (11) Any indication of AGN activity.}
\tablenotetext{a}{NIR CLs detected in \citet{Satyapal2017}.}
\end{deluxetable*}
\vspace*{-8mm}

\subsection{Extinction} \label{subsec:Extinction}

To quantify the extinction, we used the intrinsic line ratio of H$\alpha$/H$\beta$ = 3.1, typically used for AGN \citep[eg.,][]{Veilleux1987,Osterbrock1992}, and a Cardelli reddening law \citep{Cardelli1989} with an extinction factor of $R_V$ = 3.1. We used the narrow-line flux measurements from the SDSS data (see Section \ref{subsubsec:SDSS_fit}), where we have decomposed the H$\alpha$ and H$\beta$ emission into narrow and broad (outflow) components. For galaxies where no broad component is detected, we used the full emission line to obtain a flux. The value of E(B-V) for each galaxy is listed in Table \ref{tab:relations}. Note that two galaxies, J0122+0100 (Northeast region) and J0852+2928, have Balmer decrements below the intrinsic ratio so we did not apply any extinction correction to them. For the rest of this article, we use extinction corrected flux values unless otherwise specified.

\subsection{Star-formation Rates} \label{subsec:SFR}

We estimated the SFRs of our sample through the use of the [\ion{O}{2}] $\lambda$3727 doublet indicator presented in \citet{Kewley2004}. We list these measurements in Table \ref{tab:relations}. While [\ion{O}{2}] is useful for calculating SFRs, complications can arise if an AGN is present and can ionize the [\ion{O}{2}] gas in the extended emission line regions. This is mainly seen in cases where AGN ionization reaches beyond the NLR. In order to identify whether an AGN could be ionizing [\ion{O}{2}], the CL [\ion{Ne}{5}] $\lambda$3426 is often used \citep{Maddox2018}. As such, the presence of [\ion{Ne}{5}] would indicate contamination of the [\ion{O}{2}] emission from the AGN and thus it cannot be used as a SF indicator. 

We find [\ion{Ne}{5}] emission in two targets, J0929 and J1333, and thus omit their values from Table \ref{tab:relations}. For targets where we cannot confirm the absence of [\ion{Ne}{5}] (mainly due to the wavelength coverage of SDSS), we report their SFRs as upper limits.

\section{Evidence of AGN Activity} \label{sec:AGN_indicators}

\subsection{Coronal Line Detections} \label{subsec:CL_detection}

\begin{figure*}
\centering
\epsscale{1.15}
\plotone{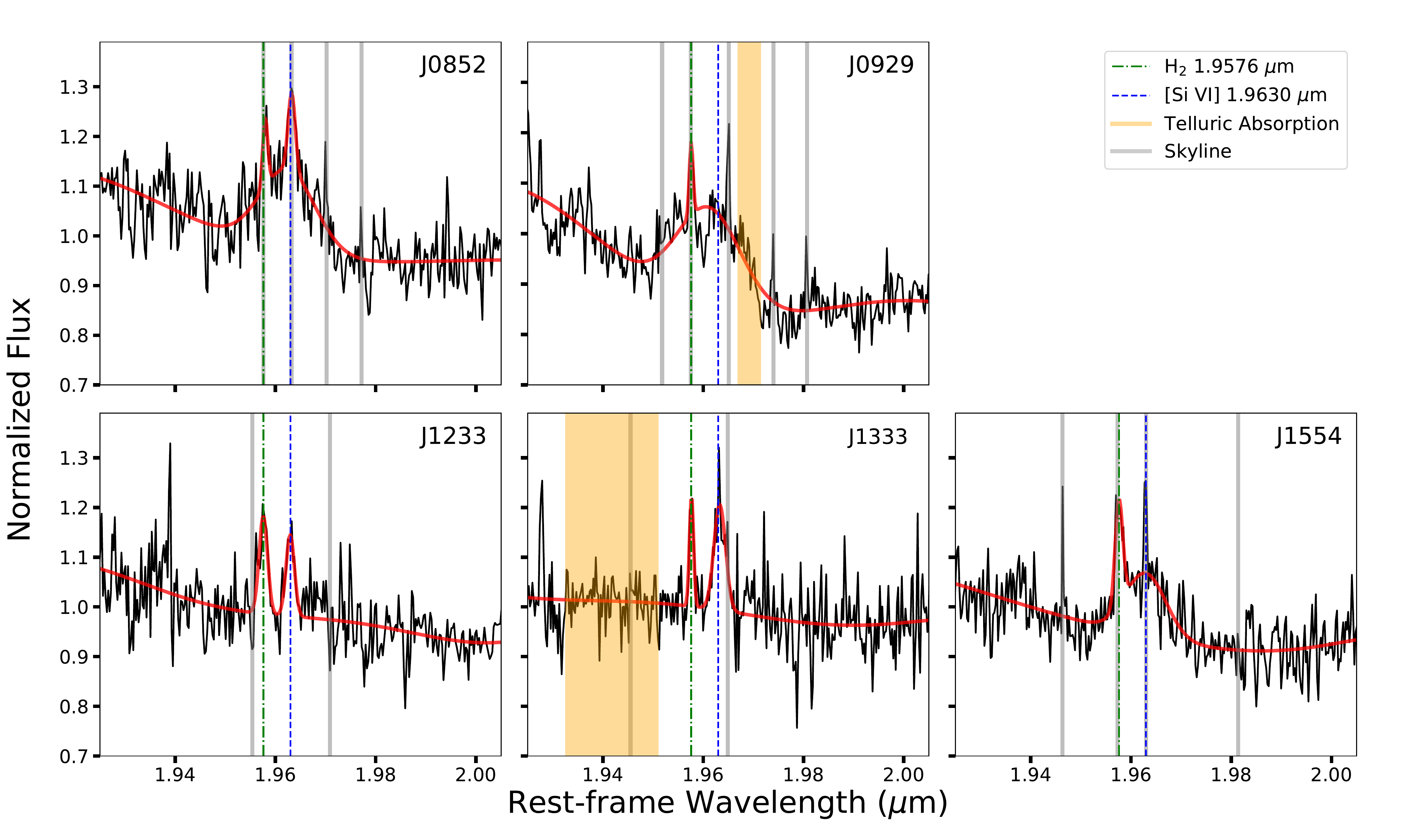}
\caption{Zoom-in plots of [\ion{Si}{6}] emission and the surrounding region. The best-fit line from \textsc{emcee} is plotted in red and the dash-dot green and blue lines denote the rest-frame location of H$_2$ 1.957 $\mu$m and [\ion{Si}{6}], respectively. Grey lines and shaded yellow regions signify the location of skylines and telluric absorption that are in close proximity to [\ion{Si}{6}]. \label{fig:CL_box}}
\end{figure*}

An effective method of identifying AGN activity is through the detection of CLs. This is because CLs have higher ionization potentials than what is typically produced by stellar processes. A total of seven out of the 13 bulgeless galaxies in our sample (54\%) show CLs.  We detect [\ion{Si}{6}] 1.9630 $\mu$m in the NIR spectra of five targets in our sample (see Figure \ref{fig:CL_box} and Table \ref{tab:relations}; fluxes listed in Table~\ref{tab:table_fluxes}).  In addition, \citet{Satyapal2017} report the detection of [\ion{Si}{6}] in J0122+0100(SE) and [\ion{Si}{10}] 1.4305 $\mu$m in J1036+0221.  They used a wider slit with a different placement for J0122+0100(SE), which may explain why we did not observe [\ion{Si}{6}] in our NIRES spectrum. We fit their [\ion{Si}{6}] emission line and include it in the following analysis. For J1036+0221, we do not detect [\ion{Si}{10}] since it is outside our selected wavelength coverage for NIRSPEC. Additional details on these two targets are included in Appendix \ref{appendix:Sample_detail}.   While we include the detection of CLs in these targets in our overall statistics, we only investigate the kinematics and other properties of the CLs observed in our NIRSPEC and NIRES spectra.

Inspection of the SDSS spectra reveals optical CL emission in only two targets, J0929+0026 and J1333+6536; these are marked with an asterisk in Table~\ref{tab:relations}. We detect [\ion{Ne}{5}] $\lambda$3426 in both of these galaxies, in addition to [\ion{Si}{6}] mentioned above.  Environmental effects have been shown to inhibit optical CL detection. As suggested by comparisons between NIR and optical spectra of Sy1 and Sy2, and the reduced strength of CLs in the latter \citep{Nagao2000,Gelbord2009}, the dusty torus likely lowers the detection frequency of such lines. This, along with the high extinction levels seen in our sample, could explain the low detection rate of optical CLs in our sample. 

\begin{figure}
\centering
\epsscale{1.1}
\plotone{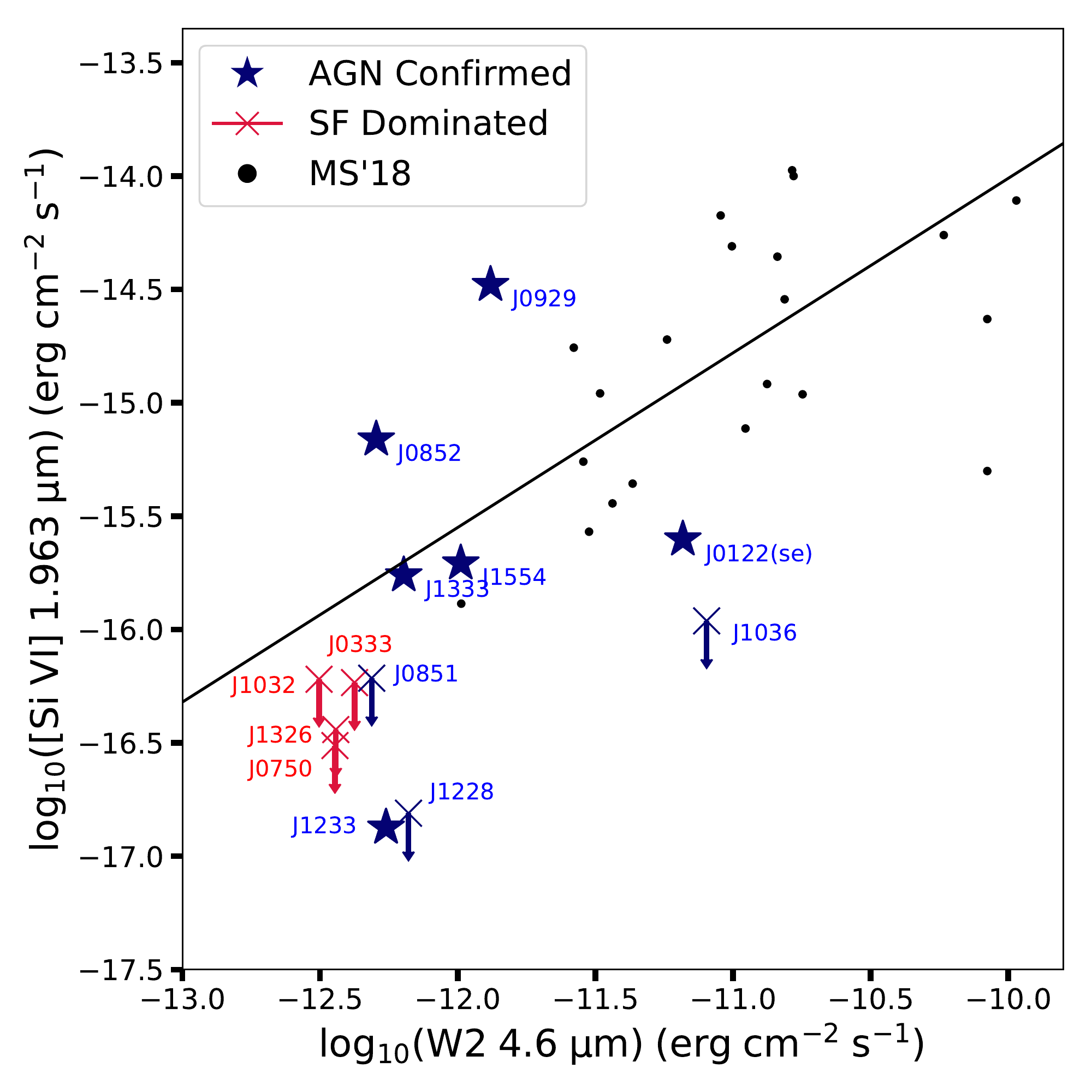}
\caption{[\ion{Si}{6}] flux vs W2 (4.6 $\mu$m) flux. [\ion{Si}{6}] detections are represented as blue stars. Note that we include the [\ion{Si}{6}] detection from S14. Crosses represent targets with no [\ion{Si}{6}] emission detected. The arrows indicate these are 3$\sigma$ upper limit fluxes to [\ion{Si}{6}]. Blue symbols represent galaxies with AGN activity while red represents galaxies with stellar emission dominating the spectra. The solid line is the best-fit line to the data found in \citet{Muller2018b}. 
\label{fig:[SiVI]_W2}}
\end{figure}

It is not surprising that [\ion{Si}{6}] is the only prominent NIR CL we detect. It has a lower ionization potential (IP = 167 eV) than other CLs and benefits from being located in a region free of stellar absorption. Indeed, other studies find [\ion{Si}{6}] to be the most common CL, with \citet{Rodriguez2011} finding it in 56$\%$ of their sample and \citet{Lamperti2017} finding it in 33$\%$ of theirs. 

The 54\% CL detection rate in our sample should be regarded as a lower limit, as there are several factors that can affect the detectability of CLs. For instance, faint lines can be easily diluted by the underlying continuum and/or they can fall below detection limits. To obtain estimates for the expected flux values for the [\ion{Si}{6}] lines that we do not detect, we make use of the relation reported by \citet{Cann2021} between \textit{WISE} \textit{W}2 (4.6 $\mu$m) flux and [\ion{Si}{6}] flux. We plot this relation in Figure \ref{fig:[SiVI]_W2}. Upper limit fluxes to the non-[\ion{Si}{6}] detections were calculated by integrating over a Gaussian with a width equaling the resolution element and the amplitude equaling the 1$\sigma$ noise level where [\ion{Si}{6}] should appear. This value was multiplied by three to obtain the 3$\sigma$ upper limit that is plotted. We include the AGN sample (black dots) from \citet{Muller2018b}, from which the plotted line of best fit is derived in order to show the scatter in the relation. Figure \ref{fig:[SiVI]_W2} shows that the majority of our non-detections (crosses) are indeed the targets with the lowest expected fluxes.

\begin{figure*}
\gridline{\fig{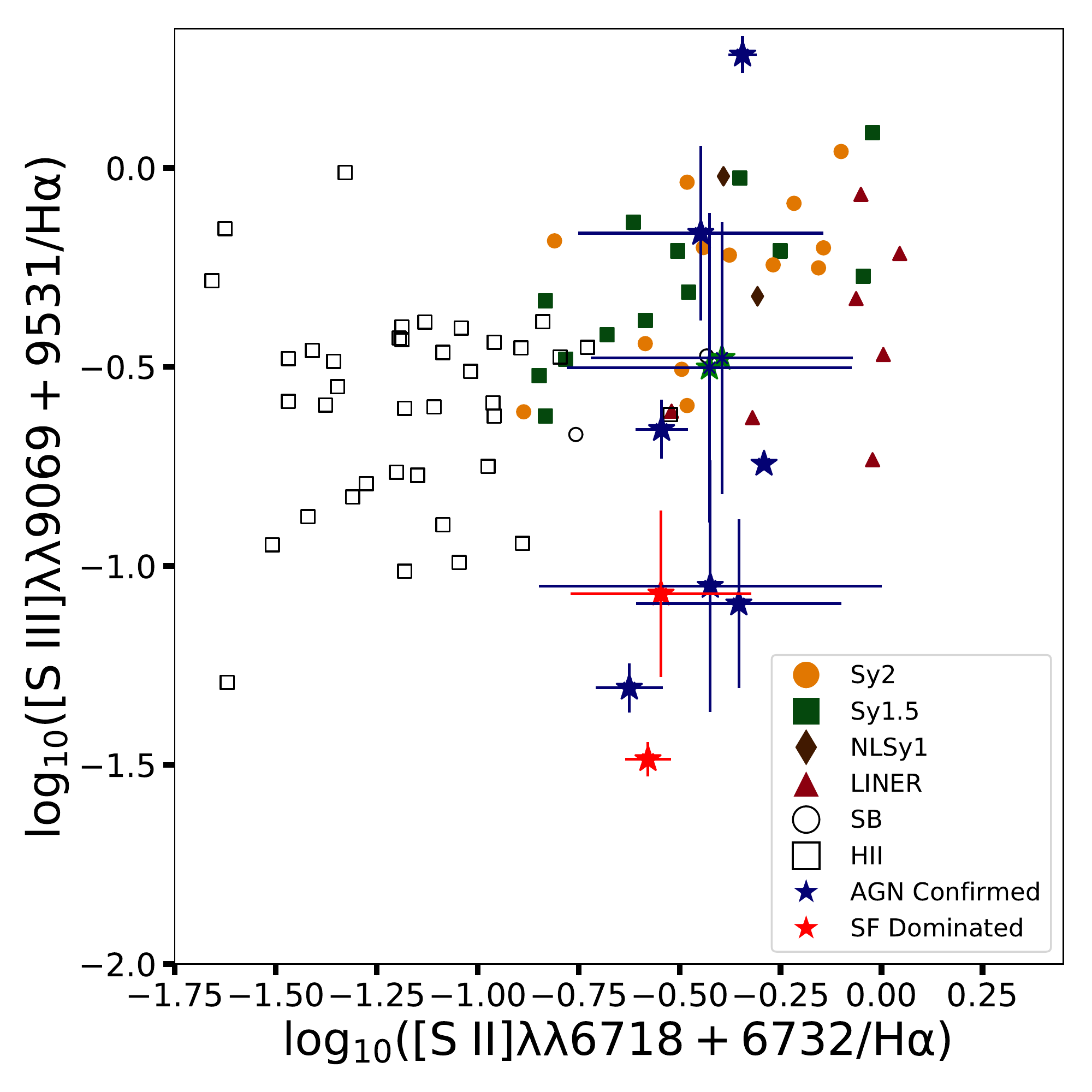}{0.34\textwidth}{}
\fig{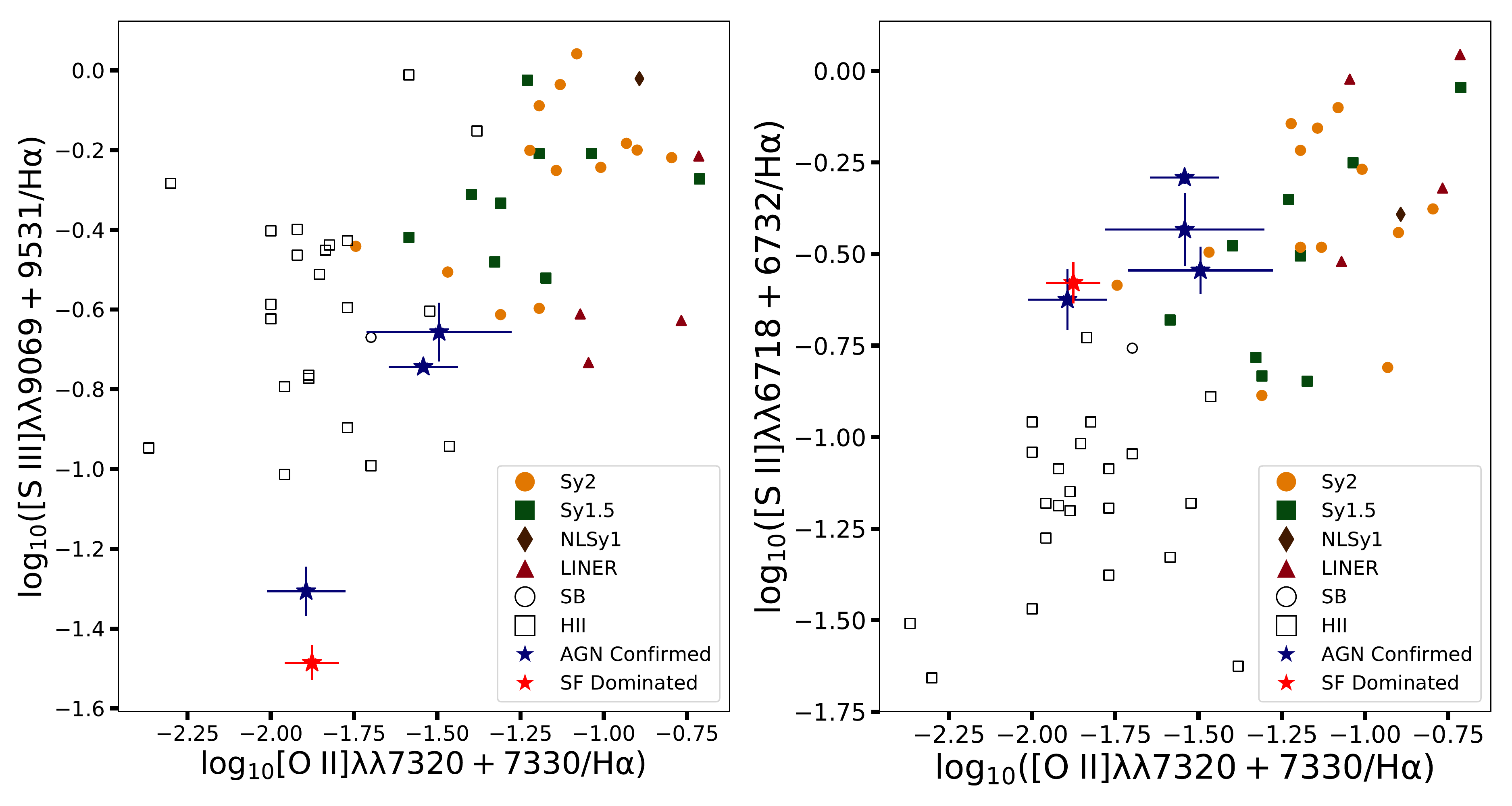}{0.64\textwidth}{}}
\caption{AGN diagnostic plots derived by \citet{Osterbrock1992}, where we include their sample. We plot our sample as blue and red stars, where blue represents galaxies with clear signatures of AGN activity and red represents galaxies whose emission is dominated by stellar processes. Fluxes to emission lines are listed in \ref{tab:table_fluxes}. \label{fig:AGN_relations}}
\end{figure*}

It is worthwhile to note the [\ion{Si}{6}] emission in J1233+0023. It has the lowest luminosity and spatial extent of [\ion{Si}{6}] in our sample. As such, this faint emission is susceptible to dilution from other sources of emission. Indeed, we see no trace of [\ion{Si}{6}] emission if we use the extraction aperture listed in Table \ref{tab:obs_log}. Only when a smaller aperture that sampled the very center was used could we confirm the presence of [\ion{Si}{6}]. Although this is the sole case in our sample, caution should be taken when using large apertures.

The majority of our non-detections lie below and to the left of the [\ion{Si}{6}] measurements, suggesting that deeper observations are likely required to detect any emission (see Figure \ref{fig:[SiVI]_W2}). Inspection of the SFRs (see Section \ref{subsec:SFR}) shows that galaxies with no [\ion{Si}{6}] emission have SFRs about an order of magnitude higher than those with [\ion{Si}{6}] emission. Visual inspection of some of these targets show that they are likely involved in an on-going merger event. In such an event, tidal forces could funnel gas towards the central regions and fuel star formation. Although these results are not indicative of starburst activity, they do suggest an increase of SF that could increase the continuum level and make weak emission lines, such as CLs, more difficult to detect.

\subsection{NIR Broad Emission from Broad-Line Region} \label{subsec:BH_mass}

The NIR spectral region includes several important recombination lines, including Pa$\alpha$ and Pa$\beta$. Like H$\alpha$ and H$\beta$, these Paschen lines can be used to trace the broad-line gas from an AGN. They also benefit from being less affected by dust obscuration. As such, the `hidden' broad-line (HBL) region can arise in cases where broad emission is present in the NIR but obscured in the optical, betraying the presence of an AGN.

Inspection of the NIR spectra reveals hidden broad-line emission (Pa$\alpha$) in two of our targets, J0851+3926 \citep[see also][]{Bohn2020} and J1228+5814. Our detection rate (15$\%$) is similar to that of other studies such as \citet{Lamperti2017}, who find HBLs in 10$\%$ of their Sy2 sample, and \citet{Veilleux1997} who find HBLs in 27$\%$ of theirs.

Observations for both of our targets with HBL were conducted on two separate nights, approximately one year apart. While broad emission is certainly indicative of AGN activity, \citet{Baldassare2016} and other follow-up studies of AGN candidates with broad emission have shown that supernovae (SNe) and other stellar activity can produce similar broad profiles. Type II SNe \citep{Pritchard2012} and luminous blue variables \citep{Smith2011} are known to produce broad recombination lines up to thousands of km s$^{-1}$. If the broad Pa$\alpha$ observed were powered by SNe, the broad emission would have persisted for more than a year, and this in turn would indicate that the SNe would likely be Type II-P. Using this time scale, we would expect to see other NIR SN features such as \ion{O}{1}, \ion{Mg}{1}, and \ion{Ca}{1} \citep[e.g.,][]{Rho2018}. However, we do not see any of these features in either of our NIR observations. Additionally, we would expect the line profiles to change significantly over this time \citep{Rho2018} but our width measurements are consistent with each other: 1490 and 1360 km s$^{-1}$ for J0851+3926, and 1640 and 1840 km s$^{-1}$ for J1228+5814.

Another potential origin of a broad-line could be powerful outflows driven by star formation. Broad, symmetric components to emission lines are observed in some starburst galaxies \citep{Martin1998,Westmoquette2008,Manzano2019}. However, these galaxies show broad emission in other lines, particularly in [\ion{O}{3}]$\lambda5007$. As mentioned in Section \ref{subsubsec:SDSS_fit}, we utilized \textsc{BADASS} to check for the presence of any outflow component in the optical spectra. All reasonable criteria came back negative for outflows in these two targets, thus it is unlikely that outflows are affecting the broad-line emission. We conclude that the most likely origin of the observed broad Pa$\alpha$ is the broad-line region of an AGN, providing strong evidence that these two targets have AGN activity.

\subsection{NIR Diagnostic Diagrams} \label{subsec:Ionize_Source}

Optical line ratios, such as the BPT diagram shown in Figure \ref{fig:WISE}, indicate AGN activity in only four of the 13 targets in our sample. With the inclusion of NIR data, we can use additional AGN diagnostics that are more insensitive to dust obscuration. \citet{Osterbrock1992} use H$\alpha$, [\ion{S}{2}] $\lambda\lambda$6718+6732, [\ion{O}{2}] $\lambda\lambda$7320+7330, and [\ion{S}{3}] $\lambda\lambda$9069+9531 fluxes to detect AGN activity. We plot these line ratio diagrams in Figure \ref{fig:AGN_relations}, where we include the AGN and star-forming samples from \citet{Osterbrock1992}. We overplot narrow-line fluxes of our sample (star symbols) and exclude the galaxies with no measurable [\ion{O}{2}] or [\ion{S}{3}] emission. In cases where [\ion{S}{3}] $\lambda$9531 is measurable but [\ion{S}{3}] $\lambda$9069 is not, we use the intrinsic flux ratio of [\ion{S}{3}] $\lambda$9531/$\lambda$9069 = 2.6 to obtain flux values for the latter. Lastly, for J0122+0100, SDSS provides three sets of spectra: one for the west galaxy and two for the east galaxy, a north and south component separated by about 2". We include all three of these in Figure \ref{fig:AGN_relations}.

Most galaxies with CLs and HBLs fall within the scatter of the other AGN. As for the rest of the sample, the line ratios of J1036+0221 and J0122+0100(ne) place them within the AGN region of Figure \ref{fig:AGN_relations}. J0122+0100(w+se) do not fall consistently in the same region but they generally do not fall within the scatter of the star-forming galaxies. In addition to the NIR line ratios presented here, \citet{Satyapal2017} report \textit{Chandra} and \textit{XMM-Newton} X-ray observations for J1036+0221 and J0122+0100(w+se). The observed luminosities of all three sources are significantly higher ($>$7.5 times greater) than the luminosities expected from X-ray binaries. This, along with CLs detected by  \citet{Satyapal2017}, indicate the presence of AGN in these galaxies.

\begin{deluxetable*}{cccccccc}
\caption{Outflow and [Si VI] Properties} 
\label{tab:outflow_kinematics}
\tablehead{\colhead{Galaxy} & \colhead{$[\rm{O\;III}]$ $v\rm{_{50}}$} & \colhead{$[\rm{Si\;VI}]$ $v\rm{_{50}}$} & \colhead{$[\rm{O\;III}]$ $W\rm{_{80}}$ ($v\rm{_{out}}$)} & \colhead{$[\rm{Si\;VI}]$ $W\rm{_{80}}$ ($v\rm{_{out}}$)} & \colhead{$[\rm{Si\;VI}]$ Radius} & \colhead{log($L_{[\rm{SiVI}]}$)} & \colhead{log($L\rm{_{AGN}}$)}\\
\colhead{ }  & \colhead{(km s$^{-1}$)} & \colhead{(km s$^{-1}$)} & \colhead{(km s$^{-1}$)} & \colhead{(km s$^{-1}$)} & \colhead{(kpc)} & \colhead{($\rm{erg\; s^{-1}}$)} & \colhead{($\rm{erg\; s^{-1}}$)}\\
\colhead{(1)} & \colhead{(2)} & \colhead{(3)} & \colhead{(4)} & \colhead{(5)} & \colhead{(6)} & \colhead{(7)} & \colhead{(8)}}
\startdata
SDSS J0852+2928 & -260$\pm$20 & -250$\pm$30 & 1120$\pm$40 (820$\pm$40) & 2130$\pm$220 (1315$\pm$140) & 0.8 & 39.97$^{+0.06}_{-0.08}$ & 42.81$^{+0.09}_{-0.12}$\\
SDDS J0929+0026 & -40$\pm$5 & -510$\pm$40 & 700$\pm$20 (390$\pm$15) & 2510$\pm$440 (1765$\pm$260) & 1.4 & 41.07$^{+0.10}_{-0.12}$ & 44.65$^{+0.06}_{-0.07}$\\
SDSS J1233+0023 & -120$\pm$20 & -70$\pm$5 & 510$\pm$60 (375$\pm$50) & 310$\pm$70 (225$\pm$40) & 0.3 & 38.19$^{+0.07}_{-0.09}$ & 42.78$^{+0.16}_{-0.25}$\\
SDSS J1333+6536 & -110$\pm$15 & -40$\pm$5 & 470$\pm$50 (345$\pm$40) & 490$\pm$50 (285$\pm$30) & 0.4 & 39.98$^{+0.04}_{-0.04}$ & 43.19$^{+0.16}_{-0.27}$\\
SDSS J1554+1457 & --- & -20$\pm$15 & --- & 1720$\pm$210 (880$\pm$120) & 1.2 & 39.99$^{+0.07}_{-0.08}$ & 42.96$^{+0.26}_{-0.35}$
\enddata
\tablecomments{Columns: (1) Galaxy name. (2) Velocity offset of the [\ion{O}{3}] $\lambda$5007 outflow component from the core component. (3) Velocity offset of [\ion{Si}{6}] from rest-wavelength. (4) $W\rm{_{80}}$ values of the [\ion{O}{3}] $\lambda$5007 outflow component. In parentheses are $v\rm{_{out}}$ values as defined in Equation \ref{eq:outflow_vel}. (5) Same as the previous column but for [\ion{Si}{6}]. (6) Spatial extension of [\ion{Si}{6}]. (7) Extinction-corrected [\ion{Si}{6}] luminosities. (8) Extinction-corrected AGN luminosities.} 
\end{deluxetable*}
\vspace{-8mm}

While no single line ratio can confidently separate AGN from stellar ionization, our bulgeless sample shows optical and NIR line ratios that, when considered collectively, strongly suggest the presence of AGN. This lack of strong stellar activity not only supports ionization from the AGN, but also indicates the outflows detected (discussed in Section \ref{sec:outflows}) are of AGN origin.

\subsection{Summary of AGN Indicators} \label{subsec:Indicator_Summary}

In the sections above, we discussed confirming AGN in our sample using a variety of methods, including CL detections, broad-line emission, and AGN diagnostic diagrams. Of our sample, we report CL detection in 54$\%$, hidden broad-lines in 15$\%$, and NIR AGN line ratios in 54$\%$. Collectively, this results in 69$\%$ (9/13) unique targets with at least one indication of AGN activity. Indeed, our results show that MIR color are an effective tool to select AGN, with over half of the confirmations coming from galaxies that would not have been classified as AGN from optical studies. For those targets for which we find no clear indication of AGN, we find they have generally twice the extinction levels and an order of magnitude higher SFRs than those for which we find evidence of AGN in the NIR. If these galaxies do indeed have an AGN, then they are likely heavily obscured or the emission from stellar processes is dominating the optical and NIR spectra. For the rest of the article, we will refer to galaxies with confirmed signatures of AGN activity as `AGN Confirmed' and those without as `SF Dominated'.

\section{Outflows} \label{sec:outflows}

In Section~\ref{subsec:Fitting}, we reported the detection of fast outflows in [\ion{O}{3}] in four of the galaxies in our sample and in Section~\ref{subsec:CL_detection} we identified an additional target with an outflow in [\ion{Si}{6}].   Thus, five out of the nine (54\%) AGN-confirmed galaxies show fast outflows, whereas none of the SF-dominated galaxies do.  This is consistent with other studies that find a higher incidence of outflows in AGN hosts as compared to star forming galaxies \citep[e.g.][W.\ Matzko et al.\ in preparation]{Cicone2014,Harrison2016,Concas2017,Leung2019,Lutz2020,Avery2021}.  

In this section, we report an observed strong correlation between the detection of outflows and that of CLs.  We also present the outflow kinematics and estimated energies, and provide supporting evidence that outflows in bulgeless galaxies may be intrinsically different than those in galaxies with merger dominated histories.

\subsection{Outflows and Coronal Lines} \label{subsec:Outflow_CL}

CLs are often associated with outflows, where the blue-shifted emission profiles of CLs have been attributed to outflow kinematics \citep{Rodriguez2006,Muller2011}. Indeed, \citet{Liu2020} and \citet{Bohn2021} find that the galaxies with the strongest and fastest outflows have the strongest and most numerous CL detections. Here, we report similar findings, where all the galaxies with detectable outflows also have CLs. We note that while J1554+1457 has NIR CL emission with a fast outflow, it does not show any evidence of an outflow in its optical spectrum. A possible reason for this is the high level of extinction, E(B-V) = 1.59, that could be inhibiting the detection of a small outflow component in the SDSS spectra.

A number of scenarios could give rise to the connection between outflows and CLs that we observe. One such case is if the AGN is responsible for both producing the outflow and ionizing the CL gas: CL emission requires a highly energetic radiative source, and this source in turn could drive a highly energetic outflow. Other scenarios are more causal. Relativistic outflowing winds have been known to produce strong shocks in the interstellar medium (ISM) surrounding the AGN \citep{Fabian2012,King2015,Harrison2018}. These shocks can heat gas, which in turn can provide at least some of the necessary ionization for CL emission. Although photoionization is believed to be the main mechanism to produce CL emission, some studies have shown that the addition of energy input from shocks can help the observed line ratios better replicate those predicted by ionization models \citep{Rodriguez2006,Muller2011}. Lastly, the outflowing winds may be simply removing the obscuring medium, allowing for a more clear view of the CL region.

\subsection{Kinematics of Outflows and Coronal Lines} \label{subsubsec:Kinematics}

Due to the connection between CLs and outflows, CLs have been used to trace the kinematics of outflows \citep{Rodriguez2006,Muller2011}. In Table \ref{tab:outflow_kinematics}, we list gas kinematics measured from the [\ion{O}{3}] and [\ion{Si}{6}] lines. To calculate the velocity of an outflow, we can utilize two values: $v_{50}$ and $W\rm{_{80}}$ \citep{Harrison2014}. Here, $v_{50}$ is the velocity offset of the outflow component from the core component while $W\rm{_{80}}$ is the width containing 80$\%$ of the line flux. For a single Gaussian profile, $W\rm{_{80}}$ = 1.09$\times$FWHM. In the cases where only one [\ion{Si}{6}] component was fit, we use the full emission profile and set $v_{50}$ to be the offset from rest-wavelength. Using these parameters, we can define the outflow velocity as follows

\begin{equation}
\label{eq:outflow_vel}
v_{\rm{out}} = -v_{50} + \frac{\rm{W_{80}}}{2}
\end{equation}

A general trend seen here is the outflow velocities associated with CLs (mean $\sim$890 km s$^{-1}$) tend to be higher than those seen in the narrow-line region (mean $\sim$480 km s$^{-1}$). Since the CL region is believed to lie closer to the SMBH, this trend would suggest a decelerating outflow. Here, a high velocity wind would originate near the AGN and would slow down as it encounters the surrounding ISM. 

We additionally report the spatial extent of [\ion{Si}{6}] emission and bolometric AGN luminosities in Table \ref{tab:outflow_kinematics}. We derived the AGN luminositites from the measured [\ion{O}{3}] $\lambda$5007 flux. Empirical bolometric correlation factors from \citet{Lamastra2009} were used: $L\rm{_{AGN}}$ = 87 $L_{[\rm{O\;III}]}$ for 38 $<$ log($L_{[\rm{O\;III}]}$) $<$ 40 and $L\rm{_{AGN}}$ = 142 $L_{[\rm{O\;III}]}$ for 40 $<$ log($L_{[\rm{O\;III}]}$) $<$ 42. We find stronger [\ion{Si}{6}] emission in galaxies with more luminous AGN, further suggesting that the AGN is the ionizing source for the CL emission in this sample. Due to the strong correlation between CLs and outflows, this would also indicate that the AGN are also responsible for driving the outflows. Although we do not find a strong correlation between outflow velocities and AGN luminosity (Pearson coefficient = -0.37), we do see a correlation with the spatial extent of [\ion{Si}{6}] and outflow speed as calculated through [\ion{Si}{6}] (Pearson coefficient = 0.86). A likely cause of this is that the faster and more energetic outflows are less inhibited by the ISM and can more easily clear out the surrounding gas. Alternatively, the faster, more extended outflows could be expanding into regions that are less dense or obscured.

\subsection{Outflow Energies} \label{subsec:outflow_energy}

The energies of the outflows can be calculated assuming a time averaged, thin-shell, free wind model \citep{Shih2010}. However, the outflows in the SDSS data are not spatially resolved and thus some simplifying assumptions need to be made on the size of the outflows. From \citet{Rupke2005}, the outflow energies can be calculated using the following simplified equation:

\begin{equation}
\label{eq:energy}
\frac{dE}{dt} = \frac{1}{2}(v_{50}^2 + 3\sigma^2)\frac{dM}{dt}
\end{equation}

\vspace{2mm}
where $\sigma$ = $W_{80}$/2.563. The outflow mass rate, $dM/dt$, can be expressed as:

\begin{equation}
\label{eq:mass_rate}
\frac{dM}{dt} = \frac{M_{out}v_{out}}{R_{out}}
\end{equation}

where $M_{out}$ is the total mass of the outflowing gas and $R_{out}$ is the upper limit on the radius of the outflow. From \citet{Osterbrock2006}, $M_{out}$ can be calculated from:

\begin{equation}
\label{eq:total_mass}
\frac{M_{out}}{M_\odot} = 4.48\bigg(\frac{L_{H\alpha}}{10^{35}\; \rm{erg}\; s^{-1}}\bigg) \bigg(\frac{<n_e>}{1000\; \rm{cm}^{-3}}\bigg)^{-1}
\end{equation}

where we assume Case B recombination with $T$ = 10$^4$ K. Here, $L_{H\alpha}$ is the extinction corrected luminosity of the outflow component of H$\alpha$ and $n_e$ is the electron density, which we estimated using the [\ion{S}{2}] $\lambda$6716/$\lambda$6731 relation \citep{Sanders2016}. The uncertainties in $n_e$ are relatively large, and are incorporated into the errors listed in Table \ref{tab:Energetics}. Lastly, upper limits to $R_{out}$ were calculated using the $R_{out}$ - $L_{[OIII]}$ relation derived in \citet{Kang2018}. We multiply these values by a factor of 2 to account for the far side of the outflow that is blocked by the galaxy. Lower limits for the mass and energy rates of the outflows are listed in Table \ref{tab:Energetics}.

\begin{deluxetable}{cccc}
\caption{Energetics of the Outflows} 
\label{tab:Energetics}
\tablehead{\colhead{Galaxy} & \colhead{log($M_{out}$)} & \colhead{log($dM/dt$)} & \colhead{log($dE/dt$)}\\
\colhead{ } & \colhead{($M_\odot$)} & \colhead{($M_\odot$ yr$^{-1}$)} & \colhead{(erg s$^{-1}$)}}
\startdata
J0852+2928 & 5.7$\pm$0.21 & $>$-0.6 & $>$40.8\\
J0929+0026 & 7.6$\pm$0.25 & $>$0.6 & $>$41.4\\
J1223+0023 & 6.1$\pm$0.24 & $>$-0.5 & $>$40.2\\
J1333+6536 & 4.3$\pm$0.19 & $>$-2.4 & $>$38.2
\enddata
\tablecomments{Columns: (1) Galaxy name. (2) Total mass of the outflow. (3) Mass outflow rate. (4) Kinetic energy outflow rate.}
\end{deluxetable}
% \vspace{3.0cm}

Comparing the mass and mass outflow rates of our galaxies to those of the disk-dominated sample from \citet{Smethurst2019}, we find that the outflows presented here have comparable values. \citet{Smethurst2019} find that the outflow rates of disk-dominated galaxies are lower than those measured in galaxies with merger dominated histories. They attribute this discrepancy to differences in black hole spin and accretion geometries that are a result of accretion by secular processes. As such, this would suggest that outflows in bulgeless galaxies may be intrinsically different than those found in galaxies that have had a significant merger history. 

\section{Scaling Relations} \label{sec:BH_Comparisons}

Since two of our targets show broad emission from the BLR, we can estimate their BH masses through the virial relation: 

\begin{equation}
\label{eq:virial}
M_{\rm{BH}} = \mathnormal{f}\frac{V^2 R}{G}
\end{equation}

where $\mathnormal{f}$ is the virial coefficient; $V$ is the velocity of the broad-line gas; $R$ is the distance from the broad emission gas to the central continuum source which can be estimated using relations derived from reverberation mapping; and $G$ is the gravitational constant. The FWHM of the broad emission line can be used for the value of $V$, and the value of $R$ is estimated empirically using the optical luminosity of the AGN as a proxy \citep{Kaspi2005,Bentz2013}.

The BH mass was obtained following the estimators presented by \citet{Kim2018}, where they adopted the virial factor log $\mathnormal{f}$ = 0.05 $\pm$ 0.12 derived by \citet{Woo2015}. From \citet{Kim2018}, their Equation 10,

\begin{equation}
\label{eq:2}
\frac{M}{M_{\odot}} = 10^{7.07\pm0.04}\left(\frac{L_{\rm{Pa}\alpha}}{10^{42} \;\rm{erg}\;\rm{s}^{-1}}\right)^{0.49\pm0.06}\left(\frac{\rm{FWHM}_{\rm{Pa}\alpha}}{10^3\;\rm{km}\;\rm{s}^{-1}}\right)^2
\end{equation}

where the FWHM of broad Pa$\alpha$ is the analog to the velocity in the virial mass estimator and $L_{\rm{Pa}\alpha}$ of the broad component is the analog to the distance to the BLR. From Equation (\ref{eq:2}), our NIRSPEC and NIRES measurements give a BH mass of log($M_{\rm{BH}}/M_{\odot}$) = 6.59$\pm$0.32 and 6.78$\pm$0.14 for J0851+3926, and 6.83$\pm$0.27 and 6.80$\pm$0.32 for J1228+5814. The listed uncertainties come from the random error estimates in the fitting process. Accounting for systematic uncertainties, virial BH mass estimates typically have errors of 0.3 - 0.5 dex \citep[eg.,][]{Shen2013,Reines2015}. For a detailed description on virial mass uncertainties, see \citet{Sexton2019}.

\citet{Bohn2020} compiled a sample of galaxies with secure BH mass estimates and obtained $M_{\rm{BH}}$--$M_{\rm{bulge}}$ and $M_{\rm{BH}}$--$M_{\rm{stellar}}$ relations. They found that bulgeless and/or disk dominated galaxies fall closer to the $M_{\rm{BH}}$--$M_{\rm{stellar}}$ relation, despite it having more scatter than the $M_{\rm{bulge}}$ relation. Moreover, they also found that bulgeless galaxies best correlate with late-type galaxies in the $M_{\rm{BH}}$--$M_{\rm{stellar}}$ relation. Indeed, with an estimated $M_{\rm{stellar}}$ = 10.61 \citep{Chang2015}, J0851+3926 only lies 0.37 dex below the late-type relation. Similarly, J1228+5814 lies 0.41 dex below the same relation.   Thus, it appears that these two bulgeless galaxies, which have presumably grown in the absence of major mergers, have managed to grow their SMBH enough to remain within the scatter of the $M_{\rm{BH}}$--$M_{\rm{stellar}}$ relation.  

\begin{figure*}
\centering
\epsscale{1.15}
\plotone{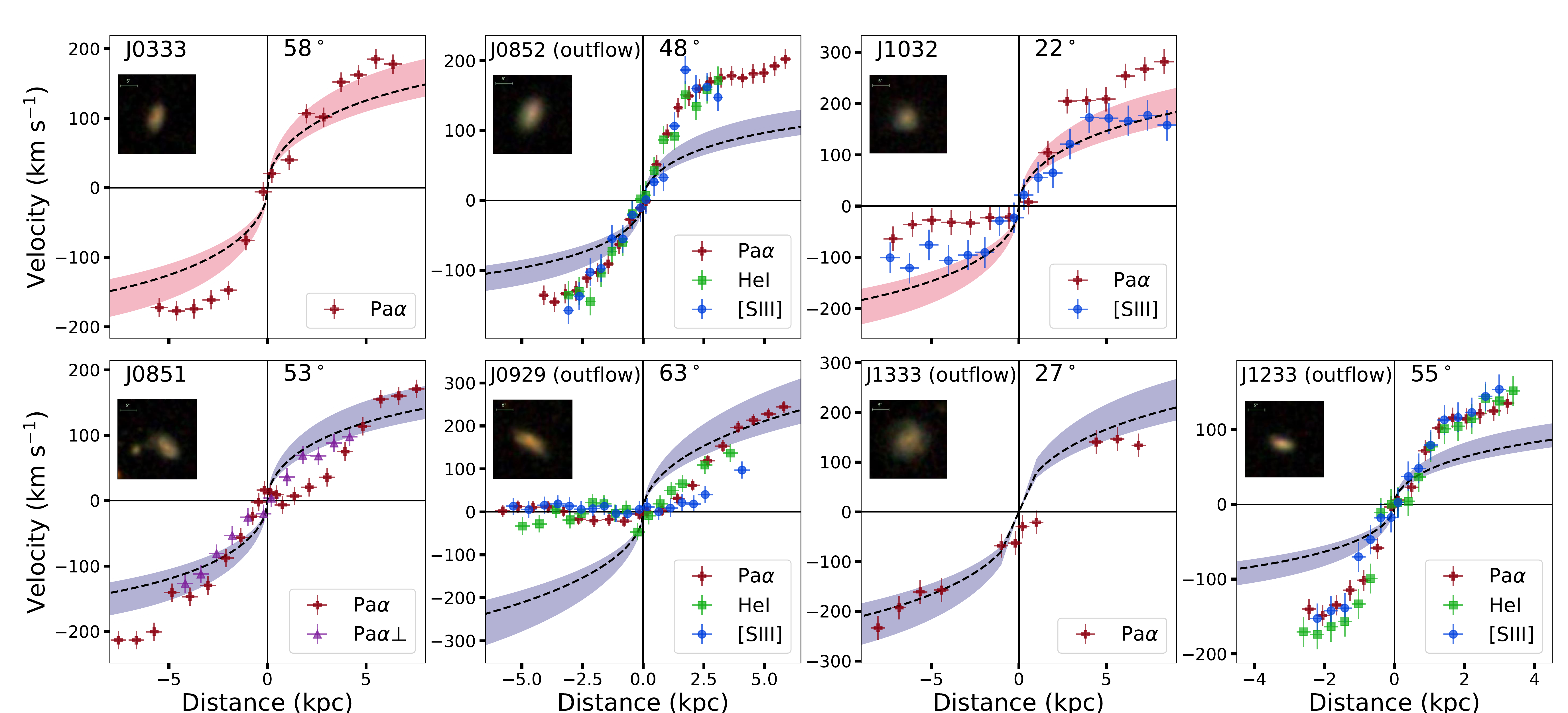}
\caption{Velocity curves, corrected for inclination, showing the gas kinematics of the galaxies in our sample that show extended emission lines. The black dotted line represents velocities expected from an NFW model, assuming a concentration parameter of 10. The shaded blue/red regions represent the concentration parameter varying from 8 to 15. Here, blue identifies targets with confirmed AGN and red identifies targets dominated by stellar emission. Images from SDSS and inclination values are also included. Notice that J0851+3926 might show signs of a counter-rotating bar, as discussed in Appendix \ref{appendix:Sample_detail} and presented in detail in \citep{Bohn2020}.
\label{fig:rotation_curve}}
\end{figure*}

An important question for AGN hosts lacking a substancial bulge is what mechanism initiated the in-fall of gas towards the center to turn the AGN on. Close inspection of J0851+3926 reveals a faint tail leading to a small galaxy (see Figure \ref{fig:Slit_pos}). We obtained spectra of this target with the goal of estimating a redshift, however the spectra is featureless in all five orders of NIRES. While it is uncertain whether this is a true interacting companion, 2-dimensional surface brightness decompostions show that J0851+3926 has undisturbed disk. Thus, it is unlikely that this event is a major merger. 
Interestingly, as we will discuss in Section \ref{sec:Gas_Kinematics}, counter-rotation is seen within the central kpc (see Figure \ref{fig:rotation_curve}). This could have been a product of an interaction with the aforementioned companion. Alternatively, a galactic bar could be present that is altering the angular momentum of the gas \citep{Piner1995,Sheth2002,Wang2012}. Follow-up, high resolution imaging will be needed to resolve the true nature of the central region.

Imaging of J1228+5814 shows two bright sources in the central region. Two-dimensional surface brightness decompositions confirm this; adding a second disk component greatly reduces the residuals of the fit. Some possible explanations of this include an off-centered AGN or an on-going merger event. Inspection of our spectroscopic data also show two sources. Only the southern one shows broad emission while the northern source only shows absorption features. There is no significant redshift difference between the absorption features in the two sources. If a merger event is indeed occurring, this could provide the in-fall of gas needed to start the gas accretion that is fueling the current AGN episode.  However, the lack of a substantial bulge still indicates a relatively mergerless past, where mergers could not have contributed significantly to the growth of the SMBH.   

In conclusion, both targets fall within the scatter of other late-type galaxies and the $M_{\rm{BH}}$ we measure are some of the highest reported for bulgeless galaxies in the literature. Although follow-up observations are needed to ascertain the triggering mechanism of the AGN, it is certainly intriguing how these BHs grew to supermassive sizes through secular processes in the absence of major mergers that would have also built up a significant bulge component.

\section{Probing Dark Matter via Rotation Curves} \label{sec:Gas_Kinematics}

Spatially extended emission is detected in a number of galaxies, enabling us to perform analysis on the kinematics of the gas (see Figure \ref{fig:rotation_curve}). We describe in Appendix \ref{appendix:vel} the procedure used to extract rotation velocities for seven targets where this was possible. The rotation curves of these galaxies showing measurable extended gas emission are shown in Figure \ref{fig:rotation_curve}. 

We mainly plot Pa$\alpha$ but also report [\ion{S}{3}] 0.953 $\mu$m and \ion{He}{1} 1.083 $\mu$m if detectable. Also note that two observed velocity profiles are plotted for J0851+3926. This is because we observed this target on two separate occasions with the slits positioned perpendicular to each other (see Figure \ref{fig:Slit_pos}). Plotted error bars include spectral and fitting errors as well as the uncertainties in the ellipticity and the value of \textit{q}.

The shaded regions of Figure \ref{fig:rotation_curve} represent velocity curves assuming a Navarro-Frenk-White (NFW) dark matter density profile \citep{Navarro1996} and using abundance matching from \citet{Moster2013} to calculate the corresponding halo mass starting from the stellar masses of our targets. We assume a concentration parameter $c = 10$ (dotted black line in Figure~\ref{fig:rotation_curve}), but vary  \textit{c} between 8 -- 15 to illustrate the uncertainty in this parameter (shaded areas). Our stellar mass, M$_\star$, estimates come from \citet{Chang2015}, who performed SED fitting on optical and IR data. These values are also consistent with the MPA/JHU catalog\footnote{\url{https://wwwmpa.mpa-garching.mpg.de/SDSS/}}. The halo mass and other associated values are listed in Table \ref{tab:rotation_parameters}.

\begin{deluxetable*}{cccccc}
\caption{Gas Rotation and Galaxy Parameters} 
\label{tab:rotation_parameters}
\tablehead{\colhead{Galaxy} & \colhead{log(M$_\star$)} & \colhead{log(M$_{\rm{DM}}$)} & \colhead{Inc. Angle} & \colhead{Outflow} & \colhead{$>$ NFW Velocities}\\
\colhead{ }  & \colhead{(M$_\odot$)} & \colhead{(M$_\odot$)} & \colhead{ } & \colhead{ } & \colhead{ }\\
\colhead{(1)} & \colhead{(2)} & \colhead{(3)} & \colhead{(4)} & \colhead{(5)} & \colhead{(6)}}
\startdata
SDSS J0333+0107 & 10.67$^{+0.10}_{-0.08}$ & 12.23$^{+0.30}_{-0.17}$ & 58$^\circ$ & --- & ---\\
SDSS J0851+3926 & 10.61$^{+0.10}_{-0.09}$ & 12.13$^{+0.26}_{-0.17}$ & 53$^\circ$ & --- & ---\\
SDSS J0852+2928 & 10.22$^{+0.09}_{-0.09}$ & 11.69$^{+0.11}_{-0.11}$ & 48$^\circ$ & \checkmark & \checkmark\\
SDDS J0929+0026 & 11.13$^{+0.02}_{-0.09}$ & 13.40$^{+0.17}_{-0.33}$ & 63$^\circ$ & \checkmark & ---\\
SDSS J1032+3617 & 10.83$^{+0.09}_{-0.10}$ & 12.56$^{+0.32}_{-0.29}$ & 22$^\circ$ & --- & ---\\
SDSS J1233+0023 & 10.01$^{+0.10}_{-0.08}$ & 11.54$^{+0.12}_{-0.10}$ & 55$^\circ$ & \checkmark & \checkmark\\
SDSS J1333+6536 & 10.93$^{+0.08}_{-0.09}$ & 12.84$^{+0.31}_{-0.30}$ & 27$^\circ$ & \checkmark & ---
\enddata
\tablecomments{Columns: (1) Galaxy name. (2) Total stellar mass of the galaxy as determined in \citet{Chang2015}. (3) Dark matter halo mass estimates as calculated through abundance matching. (4) Inclination angle of the galaxy (5) Presence of an outflow. (6) Observed rotational gas velocities are greater than NFW curve. }
\end{deluxetable*}
\vspace*{-8mm}

We find that, in general, our rotation curves are consistent with the estimates from NFW halos with masses estimated from abundance matching following \citet{Moster2013}. However, two galaxies, J0852+2928 and J1233+0023, are worth highlighting since they have observed velocities that are significantly greater than those predicted by the NFW curve. They are both isolated and thus disruptions from a merger event are unlikely. 

A more compelling, albeit speculative, interpretation is that these two galaxies inhabit a dark matter halo (DMH) that is more massive than estimates from abundance matching using their stellar mass. This would then indicate that some feedback process could have suppressed star formation, leading to the lower stellar galaxy mass. Since AGN-driven outflows are detected in both of these targets, winds could have expelled gas from the central regions and help lower star formation as a result of AGN feedback.

\begin{figure}
\centering
\epsscale{1.1}
\plotone{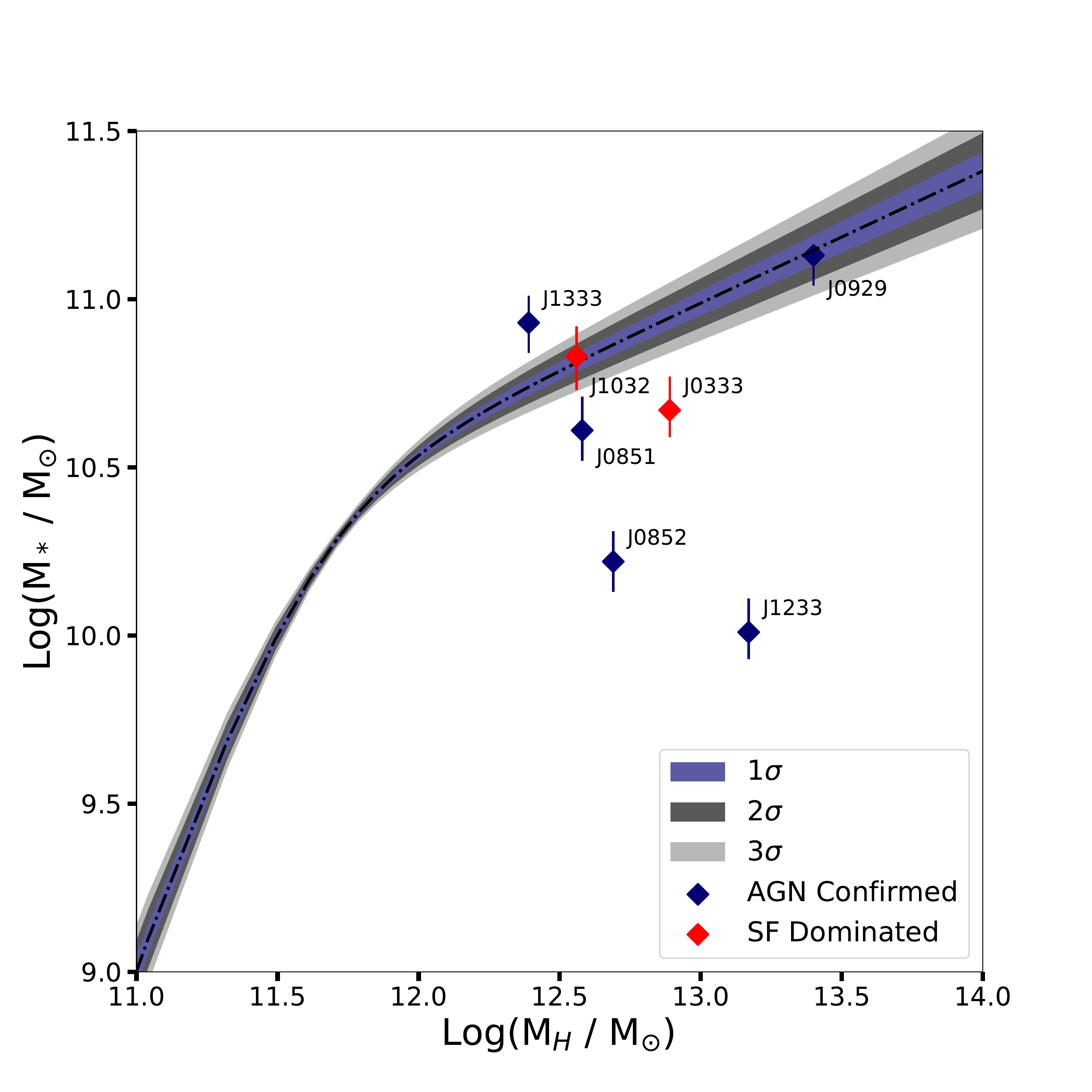}
\caption{Total stellar mass vs. halo mass relation as calculated in \citet{Moster2013} through abundance matching. The colored contours represent one, two, and three sigma errors of the relation. Plotted as diamonds are the measured stellar mass and expected DMH mass of our sample. Blue represents galaxies with confirmed AGN while red represents targets whose emission is dominated by stellar processes. \label{fig:moster}}
\end{figure}

To quantify the discrepancy between the measured stellar mass and expected DMH mass of these two targets, we plot the $M_{\rm{stellar}}$--$M_{\rm{H}}$ relation in Figure \ref{fig:moster} as derived in \citet{Moster2013}. Here, we plot the stellar mass of our sample \citep[as derived in][]{Chang2015} against the DMH mass as calculated from the observed velocities in our data. Due to their high observed velocities, the estimated DMH mass of the two targets places them about 0.7 and 1.1 dex below the relation, outside the general scatter of the other targets. 

When comparing these two with the other five targets, we see no distinguishing characteristics that would cause their stellar masses to be significantly underestimated. If the AGN is the source of regulating star formation, then this could lead to an intriguing and unique method of analyzing the effects of AGN feedback through rotation curves. However, further investigation with a larger sample and follow-up high-resolution IFS data will be needed to properly analyze the impact of AGN feedback on rotational velocities and assess the timescales on which AGN outflows affect star-formation in the local ISM.

As for the two targets with detected outflows but that do not show evidence of residing in a overmassive DMH (J0929+0026 and J1333+6536), a number of circumstances could explain why we do not see fast velocities. There is evidence of dust obscuration in J0929+0026 and a galactic bar in J1333+6536 that could be affecting the velocity profiles (see Appendix \ref{subsec:J0929} for further details). Thus, it is difficult to judge what effect, if any, the outflows may be having on the mass distribution of these galaxies.

\section{Conclusion} \label{sec:Conclusion}

In this article, we presented Keck NIR spectroscopy of a sample of candidate AGN hosted in bulgeless galaxies. These galaxies were selected through MIR selection techniques and many are not classified as AGN by optical surveys. As such, we analyzed optical and NIR spectra to confirm AGN activity, and we also examined other characteristics that may be indicative of AGN feedback. The main results are as follows:

\textbullet\; We report evidence of AGN activity in 9/13 (69$\%$) galaxies through NIR CL emission, broad emission, and \citet{Osterbrock1992} NIR line ratios. 

\textbullet\; We detect strong [\ion{Si}{6}] 1.9630 $\mu$m in five galaxies. Due to the high IP of [\ion{Si}{6}] (167 eV), this strongly suggests AGN activity. Two of these targets also have [\ion{Ne}{5}] $\lambda$3426 optical CL emission. Including detections of CL in two other targets of our sample by \citet{Satyapal2017}, our total CL detection rate is 54\% (7/13).
Most of our non-detections of CLs can be explained by emission lines falling in telluric absorption, high extinction, or requiring deeper observations. 

\textbullet\; We detect signatures of outflows in five of our targets, all of which have NIR CL detections suggesting a strong correlation between the two. The outflow velocities measured through [\ion{Si}{6}] (mean $\sim$890 km s$^{-1}$) tend to be faster than those measured through [\ion{O}{3}] (mean $\sim$480 km s$^{-1}$), which is consistent with a decelerating outflow. In addition, we see a trend between AGN luminosity and [\ion{Si}{6}] luminosity which suggests that CL emission, and thus most likely the outflows, are of AGN origin.

\textbullet\; We observe a significant correlation between the spatial extent of the [\ion{Si}{6}] emission and the velocity of the outflowing coronal gas, with the fastest outflows (W$_{80}$ $\sim$2000 km s$^{-1}$) extending to over $\sim$1 kpc from the center of the galaxy, while the slower outflows (W$_{80}$ $\sim$400 km s$^{-1}$) tend to be confined to within a few hundred pc.

\textbullet\; Estimates to the mass and energies of the outflows in our sample are similar to those of disk-dominated galaxies with AGN-driven outflows, but are less than those of bulge-dominated galaxies.   Our results are in agreement with those of \citep{Smethurst2021}, suggesting that merger-free galaxies may have outflows that are intrinsically different from those of galaxies with significant merger histories.

\textbullet\; We report virial estimates to BH masses in two galaxies of our sample. Both of these targets show broad Pa$\alpha$ with no clear indication of broad-lines in their optical spectra. Their BH mass estimates, log($M_{\rm{BH}}/M_{\odot}$) = 6.78 and 6.80, respectively, place them both well within the scatter of other late-type galaxies in the $M_{\rm{BH}}$--$M_{\rm{stellar}}$ relation. More importantly, the lack of a strong bulge component indicates that secular processes, likely independent of major mergers, grew these BHs to supermassive sizes.

\textbullet\; Inspection of rotation curves reveals two targets with observed velocities significantly greater than those predicted by the NFW curve. This would suggest an unusually massive DMH compared to their observed stellar mass. In such interpretation, some feedback process could have quenched star formation, leading to a smaller stellar mass. Since outflows likely originating from the AGN are detected in both these galaxies, AGN feedback regulating star formation is certainly an intriguing possibility.\\

\vspace{3mm}
\hspace{2cm}\textsc{ACKNOWLEDGMENTS}\\

We thank Lisa Prato and Remington Sexton for their assistance with their codes, \textsc{REDSPEC} and \textsc{BADASS}, and George Becker for assisting with the NIR reduction pipeline.

We thank Dr. Percy Gomez, Dr. Gregg Doppmann, and Dr. Josh Walawender for supporting our Keck observations. Partial support for this project was provided by the National Science Foundation, under grant No. AST 1817233. We also thank Dr. Randy Campbell and the support provided by the Keck Visiting Scholars Program.

LVS acknowledges financial support from NSF AST 1817233 and NSF CAREER 1945310 grants.

Some of the data presented herein were obtained at the W. M. Keck Observatory, which is operated as a scientific partnership among the California Institute of Technology, the University of California and the National Aeronautics and Space Administration. Use of this observatory was made possible by the generous financial support of the W. M. Keck Foundation.

The authors wish to recognize and acknowledge the very significant cultural role and reverence that the summit of Mauna Kea has always had within the indigenous Hawaiian community. We are most fortunate to have the opportunity to conduct observations from this mountain.

Funding for the SDSSI/II has been provided by the Alfred P. Sloan Foundation, the Participating Institutions, the National Science Foundation, the U.S. Department of Energy, the National Aeronautics and Space Administration, the Japanese Monbukagakusho, the Max Planck Society, and the Higher Education Funding Council for England. The SDSS Website is \url{http://www.sdss.org/}. SDSS is managed by the Astrophysical Research Consortium for the Participating  Institutions. The Participating Institutions are the American Museum of Natural History, Astrophysical Institute Potsdam, University of Basel, University of Cambridge, Case Western Reserve University, University of Chicago, Drexel University, Fermilab, The Institute for Advanced Study, the Japan Participation Group, Johns Hopkins University, The Joint Institute for Nuclear Astrophysics, the Kavli Institute for Particle Astrophysics and Cosmology, the Korean Scientist Group, the Chinese Academy of Sciences (LAMOST), Los Alamos National Laboratory, the Max Planck Institute for Astronomy (MPIA), the Max Planck Institute for Astrophysics (MPA), New Mexico State University, Ohio State University, University of Pittsburgh, University of Portsmouth, Princeton University, the United States Naval Observatory, and the University of Washington.

This publication makes use of data products from the Wide-field Infrared Survey Explorer, which is a joint project of the University of California, Los Angeles, and the Jet Propulsion Laboratory/California Institute of Technology, funded by the National Aeronautics and Space Administration.

The Pan-STARRS1 Surveys (PS1) and the PS1 public science archive have been made possible through contributions by the Institute for Astronomy, the University of Hawaii, the Pan-STARRS Project Office, the Max-Planck Society and its participating institutes, the Max Planck Institute for Astronomy, Heidelberg and the Max Planck Institute for Extraterrestrial Physics, Garching, The Johns Hopkins University, Durham University, the University of Edinburgh, the Queen's University Belfast, the Harvard-Smithsonian Center for Astrophysics, the Las Cumbres Observatory Global Telescope Network Incorporated, the National Central University of Taiwan, the Space Telescope Science Institute, the National Aeronautics and Space Administration under Grant No. NNX08AR22G issued through the Planetary Science Division of the NASA Science Mission Directorate, the National Science Foundation Grant No. AST-1238877, the University of Maryland, Eotvos Lorand University (ELTE), the Los Alamos National Laboratory, and the Gordon and Betty Moore Foundation.

\software{\textsc{BADASS} \citep{Sexton2020} \url{https://github.com/remingtonsexton/BADASS2}), \textsc{pPXF} \citep{Cappellari2004,Cappellari2017}, \textsc{PyRAF} (PyRAF is a product of the Space Telescope Science Institute, which is operated by AURA for NASA), \textsc{REDSPEC} (\url{https://www2.keck.hawaii.edu/inst/nirspec/redspec.html})}

\appendix

\section{NIR Line Fluxes}
\label{appendix:line_fluxes}

In this section, we list fluxes of the NIR emission lines detected at the 3$\sigma$ level or greater for each target. Note that fluxes are omitted if the emission has insufficient signal-to-noise, fall in heavy telluric absorption, or are outside the wavelength coverage. All fluxes are in units of 10$^{-17}$ erg s$^{-1}$.

\setcounter{table}{0}
\renewcommand{\thetable}{A\arabic{table}}

\startlongtable
\begin{deluxetable*}{ccccccccc}
\tablecaption{NIR Line Fluxes}
\label{tab:table_fluxes}
\tablehead{\colhead{Line} & \colhead{J0122ne} & \colhead{J0122se} & \colhead{J0122w} & \colhead{J0333} & \colhead{J0750} & \colhead{J0851} & \colhead{J0852} & \colhead{J0929}}
\startdata
$\rm{H}\beta$ 4862 \AA & 273 $\pm$6 & 3268 $\pm$421 & 7886 $\pm$689 & 306 $\pm$105 & 307 $\pm$162 & 144 $\pm$37 & 93 $\pm$4 & 1027 $\pm$101\\
Outflow Comp. & --- & --- & --- & --- & --- & --- & 41 $\pm$2 & 328 $\pm$35\\
$[\rm{O\;III}]$ 5007 \AA & 581 $\pm$7 & 2848 $\pm$336 & 5771 $\pm$475 & 138 $\pm$50 & 142 $\pm$72 & 155 $\pm$39 & 338 $\pm$7 & 10320 $\pm$909\\
Outflow Comp. & --- & --- & --- & --- & --- & --- & 149 $\pm$7 & 3342 $\pm$335\\
$\rm{H}\alpha$ 6565 \AA & 567 $\pm$9 & 10129 $\pm$ 935 & 24429 $\pm$1611 & 951 $\pm$210 & 951 $\pm$346 & 367 $\pm$77 & 278 $\pm$6 & 3181 $\pm$219\\
Outflow Comp. & --- & --- & --- & --- & --- & --- & 149 $\pm$6 & 1009 $\pm$82\\
$[\rm{N\;II}]$ 6585 \AA & 185 $\pm$3 & 3109 $\pm$273 & 7587 $\pm$456 & 426 $\pm$97 & 334 $\pm$124 & 204 $\pm$44 & 183 $\pm$4 & 716 $\pm$52\\
Outflow Comp. & --- & --- & --- & --- & --- & --- & 98 $\pm$2 & 228 $\pm$18\\
$[\rm{S\;II}]$ 6718 \AA & 167 $\pm$4 & 1357 $\pm$125 & 3573 $\pm$219 & 160 $\pm$39 & 184 $\pm$70 & 93 $\pm$21 & 66 $\pm$4 & 521 $\pm$39\\
Outflow Comp. & --- & --- & --- & --- & --- & --- & 36 $\pm$2 & 167 $\pm$14\\
$[\rm{S\;II}]$ 6732 \AA & 123 $\pm$3 & 1047 $\pm$96 & 2878 $\pm$181 & 127 $\pm$32 & 136 $\pm$53 & 70 $\pm$19 & 60 $\pm$3 & 387 $\pm$30\\
Outflow Comp. & --- & --- & --- & --- & --- & --- & 33 $\pm$2 & 124 $\pm$12\\
$[\rm{O\;II}]$ 7320 \AA & 11 $\pm$2 & 70 $\pm$12 & 172 $\pm$19 & --- & --- & --- & --- & 48 $\pm$15\\
$[\rm{O\;II}]$ 7330 \AA & 5 $\pm$1 & 59 $\pm$10 & 153 $\pm$18 & --- & --- & --- & --- & 54 $\pm$21\\
$[\rm{S\;III}]$ 0.9069 $\mu$m & 36 $\pm$4 & 119 $\pm$10 & 171 $\pm$11 & --- & --- & 33 $\pm$16 & 149 $\pm$35 & 152 $\pm$30\\
$[\rm{S\;III}]$ 0.9531 $\mu$m & 74 $\pm$8 & 361 $\pm$18 & 577 $\pm$19 & --- & --- & 57 $\pm$18 & 387 $\pm$23 & 550 $\pm$40\\
$\rm{Pa}\epsilon$ 0.9549 $\mu$m & --- & 27 $\pm$3 & 51 $\pm$3 & --- & --- & --- & --- & ---\\
$\rm{Pa}\delta$ 1.0052 $\mu$m & 9 $\pm$1 & 34 $\pm$4 & 58 $\pm$3 & --- & --- & --- & --- & ---\\
$\rm{He\;I}$ 1.0830 $\mu$m & --- & --- & --- & --- & --- & --- & 135 $\pm$20 & 237 $\pm$42\\
$\rm{Pa}\gamma$ 1.0941 $\mu$m & --- & --- & --- & --- & --- & --- & 20 $\pm$2 & 62 $\pm$4\\
$[\rm{P\;II}]$ 1.1880 $\mu$m & --- & --- & 12 $\pm$2 & --- & --- & --- & 36 $\pm$7 & ---\\
$[\rm{Fe\;II}]$ 1.2567 $\mu$m & 12 $\pm$0.7 & 33 $\pm$2 & 58 $\pm$3 & --- & --- & --- & 184 $\pm$12 & ---\\
$\rm{Pa}\beta$ 1.2822 $\mu$m & --- & --- & --- & --- & --- & --- & --- & 43 $\pm$3\\
$[\rm{Fe\;II}]$ 1.6435 $\mu$m & 7 $\pm$0.9 & 28 $\pm$2 & 38 $\pm$2 & --- & --- & --- & 57 $\pm$4 & ---\\
$\rm{Pa}\alpha$ 1.8756 $\mu$m & 89 $\pm$4 & 288 $\pm$9 & 514 $\pm$17 & 41 $\pm$3 & 69 $\pm$11 & 168 $\pm$18\tablenotemark{a} & 87 $\pm$3 & 427 $\pm$44\\
$\rm{Br}\delta$ 1.9451 $\mu$m & 4 $\pm$1 & 18 $\pm$1 & 27 $\pm$2 & --- & --- & --- & --- & ---\\
$\rm{H_2}$ 1.9576 $\mu$m & 3 $\pm$1 & 7 $\pm$1 & 15 $\pm$2 & --- & --- & --- & 5 $\pm$2 & 17 $\pm$4\\
$[\rm{Si\;VI}]$ 1.9630 $\mu$m & --- & --- & --- & --- & --- & --- & 69 $\pm$9 \tablenotemark{a} & 392 $\pm$102\\
$\rm{H_2}$ 2.0338 $\mu$m & 2 $\pm$0.5 & 4 $\pm$1 & 7 $\pm$2 & --- & --- & --- & --- & ---\\
$\rm{He\;I}$ 2.0587 $\mu$m & 4 $\pm$1 & 18 $\pm$2 & 21 $\pm$4 & --- & --- & --- & --- & ---\\
$\rm{H_2}$ 2.1218 $\mu$m & 4 $\pm$1 & 9 $\pm$1 & 17 $\pm$1 & --- & --- & --- & 11 $\pm$2 & ---\\
$\rm{Br}\gamma$ 2.1661 $\mu$m & 8 $\pm$1 & 31 $\pm$1 & 38 $\pm$1 & --- & --- & --- & --- & 13 $\pm$3\\
$\rm{H_2}$ 2.2235 $\mu$m & --- & 3 $\pm$0.5 & 5 $\pm$1 & --- & --- & --- & --- & ---\\
\hline \\[-1.8ex]
$W1$ 3.4 $\mu$m & --- & 21 $\pm$0.5 & --- & 2.5 $\pm$0.1 & 2 $\pm$0.1 & 3 $\pm$0.1 & 3 $\pm$0.1 & 8 $\pm$0.2\\
$W2$ 4.6 $\mu$m & --- & 66 $\pm$1 & --- & 4 $\pm$0.1 & 4$\pm$0.1 & 5 $\pm$0.2 & 5 $\pm$0.2 & 13 $\pm$0.3\\
$W3$ 12.0 $\mu$m & --- & 1028 $\pm$14 & --- & 57 $\pm$3 & 55 $\pm$3 & 44 $\pm$3 & 61 $\pm$4 & 113 $\pm$4
\enddata
\end{deluxetable*}
\addtocounter{table}{-1}
\startlongtable
\begin{deluxetable*}{cccccccc}
\tablecaption{\textit{continued}}
\tablehead{\colhead{Line} & \colhead{J1032} & \colhead{J1036} & \colhead{J1228} & \colhead{J1233 (central)} & \colhead{J1326} & \colhead{J1333} & \colhead{J1554}}
\startdata
$\rm{H}\beta$ 4862 \AA & 184 $\pm$55 & 4307 $\pm$693 & 97 $\pm$93 & 223 $\pm$78 & 90 $\pm$82 & 43 $\pm$13 & 131 $\pm$71\\
Outflow Comp. & --- & --- & --- & 25 $\pm$14 & --- & 5 $\pm$2 & ---\\
$[\rm{O\;III}]$ 5007 \AA & 197 $\pm$53 & 3597 $\pm$539 & 928 $\pm$831 & 405 $\pm$132 & 119 $\pm$105 & 202 $\pm$69 & 155 $\pm$79\\
Outflow Comp. & --- & --- & --- & 40 $\pm$22 & --- & 23 $\pm$ 10 & ---\\
$\rm{H}\alpha$ 6565 \AA & 570 $\pm$115 & 13354 $\pm$1426 & 302 $\pm$248 & 692 $\pm$163 & 277 $\pm$206 & 141 $\pm$39 & 407 $\pm$150\\
Outflow Comp. & --- & --- & --- & 86 $\pm$41 & --- & 21 $\pm$8 & ---\\
$[\rm{N\;II}]$ 6585 \AA & 228 $\pm$48 & 7517 $\pm$789 & 174 $\pm$145 & 419 $\pm$108 & 275 $\pm$204 & 109 $\pm$30 & 307 $\pm$113\\
Outflow Comp. & --- & --- & --- & 46 $\pm$23 & --- & 16 $\pm$6 & ---\\
$[\rm{S\;II}]$ 6718 \AA & 94 $\pm$25 & 2764 $\pm$311 & 88 $\pm$75 & 177 $\pm$46 & 103 $\pm$82 & 25 $\pm$8 & 89 $\pm$39\\
Outflow Comp. & --- & --- & --- & 20 $\pm$9 & --- & 4 $\pm$1 & ---\\
$[\rm{S\;II}]$ 6732 \AA & 68 $\pm$17 & 2160 $\pm$250 & 25 $\pm$23 & 131 $\pm$35 & 45 $\pm$39 & 25 $\pm$9 & 64 $\pm$30\\
Outflow Comp. & --- & --- & --- & 15 $\pm$7 & --- & 4 $\pm$1 & ---\\
$[\rm{O\;II}]$ 7320 \AA & --- & 240 $\pm$68 & --- & --- & --- & --- & ---\\
$[\rm{O\;II}]$ 7330 \AA & --- & 144 $\pm$69 & --- & --- & --- & --- & ---\\
$[\rm{S\;III}]$ 0.9069 $\mu$m & 14 $\pm$6 & --- & 37 $\pm$28 & 16 $\pm$5 & --- & 32 $\pm$7 & 6 $\pm$2\\
$[\rm{S\;III}]$ 0.9531 $\mu$m & 34 $\pm$6 & --- & 58 $\pm$39 & 39 $\pm$6 & --- & 65 $\pm$12 & 26 $\pm$8\\
$\rm{Pa}\epsilon$ 0.9549 $\mu$m & --- & --- & --- & --- & --- & --- & ---\\
$\rm{He\;I}$ 1.0830 $\mu$m & 32 $\pm$7 & --- & 78 $\pm$40 & 12 $\pm$2 & --- & 46 $\pm$6 & 22 $\pm$7\\
$\rm{Pa}\gamma$ 1.0941 $\mu$m & 6 $\pm$1 & --- & 15 $\pm$9 & 3 $\pm$1 & --- & --- & ---\\
$[\rm{P\;II}]$ 1.1880 $\mu$m & --- & --- & --- & --- & --- & --- & ---\\
$[\rm{Fe\;II}]$ 1.2567 $\mu$m & --- & --- & --- & 4 $\pm$1 & --- & --- & ---\\
$\rm{Pa}\beta$ 1.2822 $\mu$m & --- & --- & 89 $\pm$50 & --- & --- & 8 $\pm$3 & 8 $\pm$3\\
$[\rm{Fe\;II}]$ 1.6435 $\mu$m & --- & --- & --- & 3 $\pm$1 (2 $\pm$0.6) & --- & --- & ---\\
$\rm{Pa}\alpha$ 1.8756 $\mu$m & 33 $\pm$3 & --- & 125 $\pm$42\tablenotemark{a} & 19 $\pm$2 & 72 $\pm$20 & 46 $\pm$3 & 43 $\pm$5\\
$\rm{Br}\delta$ 1.9451 $\mu$m & --- & 345 $\pm$19 & --- & --- & --- & --- & ---\\
$\rm{H_2}$ 1.9576 $\mu$m & --- & 987 $\pm$56 & --- & 3 $\pm$1 & 39 $\pm$14 & 7 $\pm$1 & 7 $\pm$2\\
$[\rm{Si\;VI}]$ 1.9630 $\mu$m & --- & --- & --- & (2 $\pm$0.3) & --- & 18 $\pm$2 & 24 $\pm$6\\
$\rm{H_2}$ 2.0338 $\mu$m & --- & 398 $\pm$19 & --- & --- & --- & --- & ---\\
$\rm{He\;I}$ 2.0587 $\mu$m & --- & 393 $\pm$19 & --- & --- & --- & --- & ---\\
$\rm{H_2}$ 2.1218 $\mu$m & --- & 1049 $\pm$29 & --- & 3 $\pm$0.6 (1 $\pm$0.3) & --- & --- & 4 $\pm$1\\
$\rm{Br}\gamma$ 2.1661 $\mu$m & --- & 714 $\pm$25 & --- & 2 $\pm$0.8 (2 $\pm$0.6) & --- & --- & ---\\
$\rm{H_2}$ 2.2235 $\mu$m & --- & 310 $\pm$16 & --- & --- & --- & --- & ---\\
\hline \\[-1.8ex]
$W1$ 3.4 $\mu$m & 2 $\pm$0.1 & 32 $\pm$1 & 5 $\pm$0.1 & 3 $\pm$0.1 & 4 $\pm$0.1 & 5 $\pm$0.1\\
$W2$ 4.6 $\mu$m & 3 $\pm$0.1 & 80 $\pm$2 & 7 $\pm$0.2 & 5 $\pm$0.2 & 6 $\pm$0.2 & 10 $\pm$0.2\\
$W3$ 12.0 $\mu$m & 43 $\pm$3 & 1503 $\pm$31 & 28 $\pm$2 & 73 $\pm$3 & 64 $\pm$2 & 125 $\pm$4
\enddata
\tablecomments{All optical and NIR fluxes are in units of 10$^{-17}$ erg s$^{-1}$. \textit{WISE} fluxes are in units of 10$^{-13}$ erg s$^{-1}$. J1036 also has emission lines with the following fluxes: $\rm{H_2}$ 2.0735 $\mu$m (106 $\pm$14) and $\rm{H_2}$ 2.2477 $\mu$m (182 $\pm$15).}
\tablenotetext{a}{Includes both broad and narrow component fluxes.}
\end{deluxetable*}
\vspace{-8mm}
\section{Additional Details on Individual Objects}
\label{appendix:Sample_detail}

\subsection{J012218.11+010025.76} \label{subsec:J0122}

J0122+0100 is a dual system that is likely undergoing a merger event. Optical and NIR spectroscopy identify three sources, two in the east component (NE and SE) and one in the west. The two east sources are separated about 1.9" (2.1 kpc). Optical line ratios place all three sources in the star-forming region of the BPT diagram. NIR line ratios of the southeast and west sources, likely the nuclei of the two galaxies, also lie within the scatter of star-forming galaxies in the NIR diagnostic diagrams of \citet{Osterbrock1992}. Indeed, these sources show prominent \ion{H}{1} recombination, \ion{He}{1}, and H$_2$ emission lines. In Pa$\alpha$, spatially extended gas emission is seen in the East component. It extends northwards by about 3.65" (3.9 kpc) and has receding velocities. Due to the asymmetry of the emission, it is likely a result of the interaction between the two galaxies. 

While we do not detect any CLs in any of the three sources in our observations, \citet{Satyapal2017} report a 4$\sigma$ detection of [\ion{Si}{6}] in the southeast source based on LBT LUCI-1 spectroscopy. Close inspection of the acquisition image reveals three sub-components in the eastern target. Due to the smaller width of the NIRES slit (0.55", compared to 1.0" of LUCI), we could have missed this third component since it is not aligned with the other two ($\sim$0.4" offset). 

\textit{Chandra} and \textit{XMM-Newton} X-ray observations are also reported  for both the east and west sources \citep{Satyapal2017,Pfeifle2019}. The observed X-ray luminosities are above those expected by stellar processes, such as XRBs. Coupled with the high extinction found in these targets, these results point to an optically obscured dual AGN system.

\subsection{J033331.86+010716.92} \label{subsec:J0333}

J0333+0107 was observed under heavy cloud cover and only $K$-band data have sufficient S/N for proper analysis. As such, this could severely effect the detections of any weak CLs. The high extinction, E(B-V) = 1.38, could also play a role in the lack of CL detections. Our slit position for this target is misaligned $\sim$15$^\circ$ from the semi-major axis. As such, the rotational velocities in Figure \ref{fig:rotation_curve} could be underestimated.

\subsection{J085153.64+392611.76} \label{subsec:J0851}

J0851+3926 is one of two galaxies for which we report a BH mass through virial mass estimators using broad Pa$\alpha$ (see Section \ref{subsec:BH_mass} and \ref{sec:BH_Comparisons}). Here, the broad Pa$\alpha$ is detected while no clear indication of optical broad lines are seen in the SDSS spectra. We can use this broad emission (or lack thereof) to calculate estimates to extinction in the central regions. To do this, we fit the SDSS spectrum incorporating a broad component to H$\alpha$. A check of the \textit{F}-test (\textit{F} = $(\sigma_{single})^{2}/(\sigma_{double})^{2}$) results in \textit{F} = 1.41, too low to justify adding the component. As such, we will consider these measurements as upper limits. Fits to the spectra give a line ratio of Pa$\alpha$/H$\alpha$ $\geq$ 1.44. Assuming an electron density of $n_e$ = 10$^8$ cm$^{-3}$ and temperature $T_e$ = 15,000 K, we use the intrinsic line ratio Pa$\alpha$/H$\alpha$ = 0.10 as derived in \citet{Dopita2003}. This results in an E(B-V) $\geq$ 1.40, higher than the extinction measured using the H$\alpha$/H$\beta$ Balmer decrement. This extinction is comparable to that of other targets that do not show any CLs, suggesting that obscuration could be inhibiting CL detection in this target.

As discussed in Section \ref{sec:Gas_Kinematics}, inspection of gas emission in J0851+3926 shows counter-rotation within the central kpc. This asymmetry is not seen in the slit perpendicular to this, which could indicate the presence of a bar along the semi-major axis \citep{Bohn2020}. Analysis of SDSS imaging seems to confirm this, but higher resolution imaging is required. Another scenario to create the observed counter-rotation is through tidal disruptions induced by a flyby of a companion galaxy (see SDSS postage stamp in Figure \ref{fig:rotation_curve}). However, the spectroscopic redshift of this companion could not be obtained and its photometric redshift does not match with J0851+3926. If the photometric redshifts are accurate, this would indicate the companion is a background galaxy. Surface brightness decompositions of J0851+3926 seems to confirm this, since no significant asymmetries in the disk appear and thus it is largely intact. Regardless of the scenario, it is likely through these processes that gas can be funneled to the central regions to allow BH growth without the need of major mergers or the build-up of the central bulge.

\subsection{J092907.78+002637.29} \label{subsec:J0929}

The velocity profile of the gas in J092907.78+002637.29 is quite unique since it is significantly asymmetric. We see receding gas that roughly matches the NFW curve, though it is somewhat `core-like'. We see zero net velocity in the gas emission on the approaching side. A number of factors could be the cause of this, but due to the low resolution of available imaging, we can only provide speculative scenarios. The simplest reasoning is that the rotating gas did not fall within the slit, however this is unlikely since the slit covers most of the semi-major axis. A more plausible scenario is that a dust plane is obscuring the rotating gas and that we are only observing gas moving perpendicular to us. This is consistent with the extinction seen in this target, E(B-V) = 1.28, one of the highest of targets with rotation curves. 

Considering the asymmetry in the profile, the possibility of the outflow or a merger event altering the gas rotation cannot be dismissed. While we detect a fast outflow in J0929+0026, the associated gas emission is blueshifted and is inconsistent with the rotation curve. If we assume a bi-conical model \citep{Bae2016}, we would also expect the receding gas to be more heavily obscured by the galactic disk. Lastly, while surface brightness decompositions do not show a significant merger event occurring, higher resolution imaging will be needed to confirm if a minor merger is occurring.

\subsection{J103631.88+022144.10} \label{subsec:J1036}

HST imaging reveals J1036+0221 is likely a merger system. Our NIR data show rotational H$_2$ (2.0338, 2.1218 $\mu$m), \ion{He}{1} 2.0587 $\mu$m, and Br$\gamma$ gas emission within the central $\sim$2 kpc that peaks $\sim$ 200 km s$^{-1}$. We suspect that this rotation is due largely to tidal disruptions of the merger.

Our NIRSPEC spectra, in which only $K$-band is observed, do not show any CLs. However, \citet{Satyapal2017} obtained LBT LUCI-1 spectra, through which the coverage is extended to $H$-band. They detect [\ion{Si}{10}] 1.4305 $\mu$m at the 4$\sigma$ level. NIR line ratio diagnostics (see Section \ref{subsec:Ionize_Source}) places this galaxy within the scatter of other AGN, further suggesting the presence of an AGN in the nucleus. They also detect a strong X-ray source, which further indicates AGN activity in this merger.

\subsection{J122809.19+581431.40} \label{subsec:J1228}

SDSS imaging shows evidence of two central sources in J1228+5814. This is confirmed with \textsc{GALFIT} where two exponential disks best fit the surface brightness profile. Indeed, two sources are also seen in the 2D spectrum. However, we only detect hidden broad Pa$\alpha$ emission in the south source (see Section \ref{subsec:BH_mass}). Inspection of the north source reveals a featureless spectrum, except for a few stellar absorption features in $H$-band. This is, however, enough to confirm that there is no significant velocity offset between the two sources. 

It is interesting to note that we may be observing an off-nuclear BH in J1228. The southern source is about 1.7 kpc offset from the photometric center. Simulations have shown that off-center or ``wandering" BHs can be found a few kpc away from the center in dwarf galaxies \citep{Bellovary2019}. These wandering BHs have also been found in observations of dwarf galaxies \citep{Reines2020} but are not as common in more massive galaxies, such as J1228 (log(M$_\star$) = 10.67). As such, J1228 could lead to an interesting study of off-center AGN in more massive galaxies.

Along with J1326+1014, this target also has one of the highest extinction levels in this sample, E(B-V) = 2.50. This is largely due to the weak H$\beta$ emission that indicates significant obscuration. For J1228+5814, this is consistent with the fact that no broad H$\alpha$ emission is detected in the SDSS spectra. This high level of obscuration is likely inhibiting CL detection.\\

\section{Rotation Curves Extraction} \label{appendix:vel}
We obtained velocities at various radii by first fitting a Gaussian along the spatial axis of the rectified 2D spectrum to locate the center of the galaxy. The row with the highest pixel value was taken as the center and used as the zero-point from which we constructed velocity curves. If multiple rows in the central region had comparable pixel values, we took the pixel pair with the highest values as the center. Note that this method selects the photometric peak and not necessarily the kinematic center of the galaxy.

Spectra were extracted at 0.3" intervals (based on the telescope resolution) but since the typical seeing was worse than this, the effective extraction intervals are closer to 0.5". These intervals were increased if the S/N was sufficiently low (S/N $<$ 10). Each subsequent extraction sampled a new set of pixels out to the edge where a proper S/N of the emission line was no longer achievable. Each extraction was fit with \textsc{emcee} in a similar manner as described in Section \ref{subsubsec:NIR_fit}. The peaks of each emission profile was compared to the zero-point to obtain line of sight velocities. We convert these velocities to rotational velocities using 

\begin{equation}
\label{eq:3}
v_{\rm{rot}} = \frac{v_{\rm{LoS}}}{\rm{sin}^{-1}(\theta_{\rm{inc}})}
\end{equation}

where $\theta_{\rm{inc}}$ is the inclination angle of the disk to the line of sight. For an infinitely thin disk, $\theta_{\rm{inc}}$ can be calculated using the ratio of the semi-minor axis \textit{b} and semi-major axis \textit{a} through the equation $\theta_{\rm{inc}}$ = cos$^{-1}$(\textit{b}/\textit{a}). To more accurately estimate the inclination angle, a thickness parameter \textit{q} = \textit{c}/\textit{a} can be introduced, where \textit{a} is the radius of a circular disc and \textit{c} is its thickness. When projected on a plane, the resulting shape is elliptical and $\theta_{\rm{inc}}$ can be described as

\begin{equation}
\label{eq:4}
\theta_{\rm{inc}} = \sqrt{{\rm{cos}}^{-1}\bigg(\frac{(b/a)^2 - q^2}{1 - q^2}\bigg)}
\end{equation}

We assume a value of \textit{q} = 0.1 \citep{Giovanelli1994} for all of our targets. To calculate the ellipticity $\epsilon$ = (1 - (\textit{b}/\textit{a})), we use the \textsc{isophote} routine of the \textsc{photutils} package \citep{Bradley2020}.

Seven galaxies show spatially extended emission for which we could measure the rotation curves as shown in Figure~\ref{fig:rotation_curve}. We also do see some extended gas emission in J1036+0221 and J0122+0100(w), but do not include them in this analysis since they are likely involved in a merger event. The remaining four targets do not show spatially extended emission lines, most likely due to the high levels of extinction that they show.

\section{J120443.76+050543.87}
\label{appendix:J1204}

We include J1204+0505 here because it was observed in a parallel campaign and shares many similar characteristic with the bulgeless sample presented here. It is located at a redshift of 0.1097 and has a stellar mass of 10.1 M$_\odot$ \citep{Chang2015}. Surface brightness decompositions from \citet{Simard2011} show a B/T = 0.01, and supplementary fits only using an exponential disk result in a S{\'e}rsic index of 0.71. As such, it is unlikely that there is a significant bulge component. Although the BPT optical line ratios are not suggestive of AGN activity, J1204+0505 does fall within the scatter of our sample in the [\ion{S}{3}]/H$\alpha$ vs. [\ion{S}{2}]/H$\alpha$ diagram. It also meets the MIR color cut of S14 ($W1$ -- $W2$ = 0.84).

\begin{figure*}
\centering
\epsscale{1.15}
\plotone{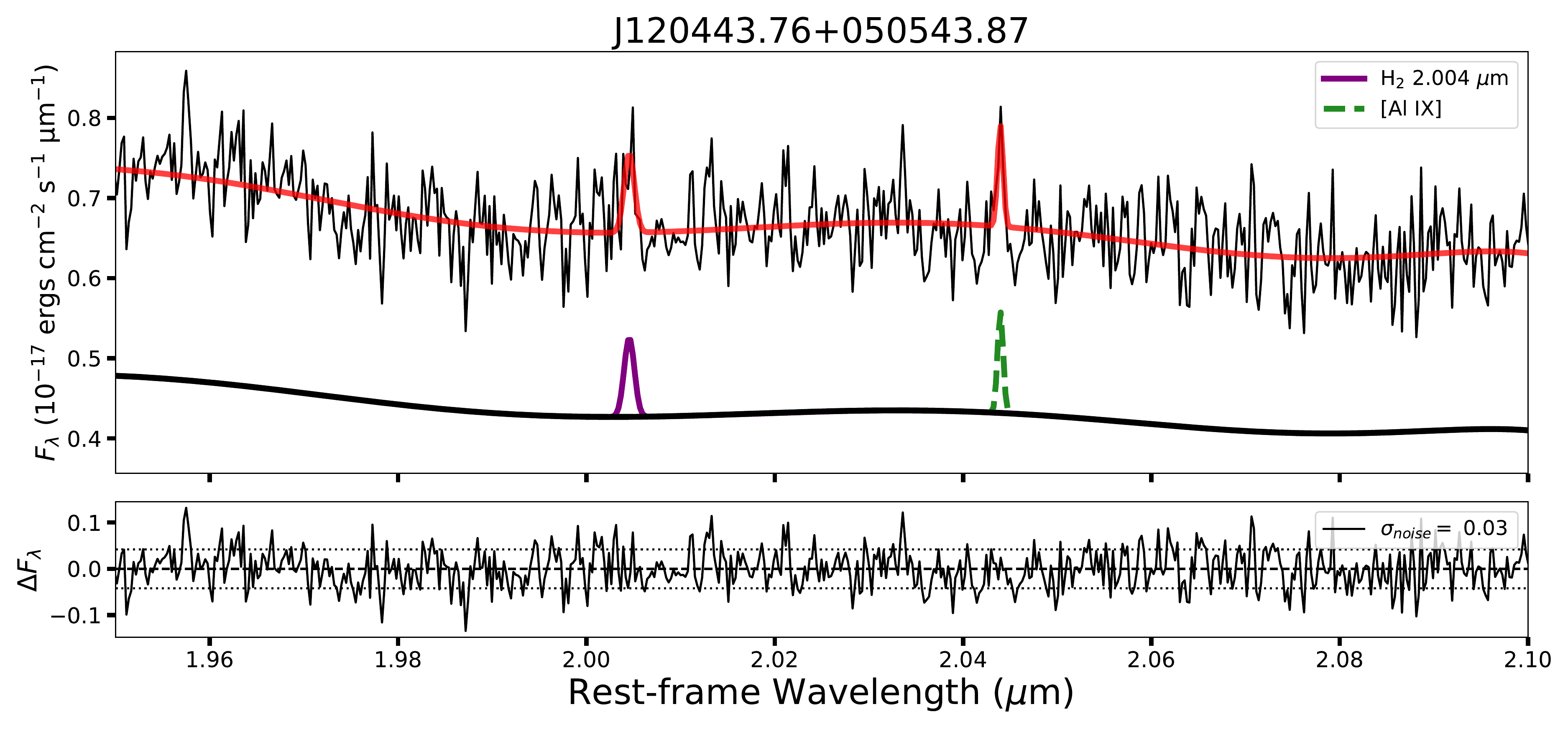}
\caption{Zoom-in plot of [\ion{Al}{9}] 2.044 $\mu$m emission in J1204. The best-fit line from \textsc{EMCEE} is plotted in red. The fit to H$_2$ 2.004 $\mu$m is also shown. The bottom panel shows the residuals of the fit with the 1$\sigma$ error level shown as horizontal dotted lines. \label{fig:J1204}}
\end{figure*}

J1204+0505 is a unique case since it is the only galaxy where we detect [\ion{Al}{9}] 2.044 $\mu$m emission and no other CL. Fitting the line with a single Gaussian, we detect it at a 3$\sigma$ level (see Figure \ref{fig:J1204}). To determine whether noise could have recreated this line profile, we use the \textit{F}-test in a regression analysis model. Put simply, we sample a small region around the emission line and compare the residuals of when the line is fit and when it is not. We will call this value $F_{statistic}$ (or $F_{stat}$). Since this value is based on residual error, we need to treat it as a random variable with some probability. As such, we can use this to construct a probability density function (PDF), from which we can obtain a confidence level that the emission line is real, and not from random noise.

We begin by using the following equation to calculate $F_{stat}$:

\begin{equation}
\label{eq:total_mass}
\textit{F$_{stat}$} = \bigg(\frac{\frac{RSS_1 - RSS_2}{k_2 - k_1}}{\frac{RSS_2}{n - k_2}}\bigg) 
\end{equation}

where $RSS_1$ is the residual sum of squares of just a continuum fit (one degree of freedom), $RSS_2$ is the residual sum of squares of a continuum plus Gaussian fit (four degrees of freedom), and $k$ is the degrees of freedom for each fit. Here, $n$ is the sampling size, which in this case, is the number of wavelength channels in the selected region. From here, ($k_2 - k_1$) and ($n - k_2$) can be used to calculate the PDF for $F_{stat}$. We obtain a probability value of 0.97, that is, we detect the line at a 97$\%$ confidence. While this result is suggestive that the emission line is real, deeper follow-up observations will be needed to confirm this.

\bibliography{Paper.bib}{}
\bibliographystyle{aasjournal}

\end{document}